# Modeling Stock Price Dynamics with Fuzzy Opinion Networks

Li-Xin Wang
Department of Automation Science and Technology
Xian Jiaotong University, Xian, P.R. China
Email: lxwang@mail.xjtu.edu.cn

## Abstract

We propose a mathematical model for the word-of-mouth communications among stock investors through social networks and explore how the changes of the investors' social networks influence the stock price dynamics and vice versa. An investor is modeled as a Gaussian fuzzy set (a fuzzy opinion) with the center and standard deviation as inputs and the fuzzy set itself as output. Investors are connected in the following fashion: the center input of an investor is taken as the average of the neighbors' outputs, where two investors are neighbors if their fuzzy opinions are close enough to each other, and the standard deviation (uncertainty) input is taken with local, global or external reference schemes to model different scenarios of how investors define uncertainties. The centers and standard deviations of the fuzzy opinions are the expected prices and their uncertainties, respectively, that are used as inputs to the price dynamic equation. We prove that with the local reference scheme the investors converge to different groups in finite time, while with the global or external reference schemes all investors converge to a consensus within finite time and the consensus may change with time in the external reference case. We show how to model trend followers, contrarians and manipulators within this mathematical framework and prove that the biggest enemy of a manipulator is the other manipulators. We perform Monte Carlo simulations to show how the model parameters influence the price dynamics, and we apply a modified version of the model to the daily closing prices of fifteen top banking and real estate stocks in Hong Kong for the recent two years from Dec. 5, 2013 to Dec. 4, 2015 and discover that a sharp increase of the combined uncertainty is a reliable signal to predict the reversal of the current price trend.

# I. Introduction

Stock price dynamics are mainly driven by the expectations of investors: If an investor expects that the price of a stock will increase (the current price undervalues the stock), then he (she) will buy the stock, causing the price of the stock to rise; in the other direction, if the investor expects that the price will decline (the current price overvalues the stock) and therefore sells the stock, the price will decrease. It is the interplay between these buying and selling forces that drives the dynamics of the stock prices. How do investors form their expectations about the stock prices? This is one of the most important and fundamental questions in financial economics, and a number of Nobel Prizes were awarded to the researches related to this question, such as the 1990 Nobel Prize of Economics to the CAPM (Capital Asset Pricing Model) of Sharpe [42] and the 2013 Nobel Prize of Economics to Fama's Efficient Market Hypothesis [15] and Factor Models [17] and to Shiller's behavioral finance approach to stock pricing [44] which emphasizes human psychology (such as animal spirit [34]) as the major driving force for stock price dynamics.

The basic stock pricing model in classic financial economic theory is the following dividend-discounting model [14]:

$$\bar{p}_t = E_t \sum_{k=0}^{\infty} \frac{D_{t+k}}{(1+r)^{k+1}} \tag{1}$$

where $\bar{p}_t$ is the expected price of the stock at time $t$, $D_{t+k}$ is the dividend payout for the stock at time $t + k$, $r$ is discounting rate representing the time-value of money, and $E_t$ is the expectation operator conditional on all publicly available information about the stock at time $t$. Model (1) says that the expected price of a stock equals the summation of all the future dividend payouts discounted by the rate $r$. Although providing a good theoretical foundation, model (1) is difficult to use in practice because there are no reliable ways to determine the future dividends $D_{t+k}$ at current time $t$.

Another benchmark model for stock pricing is Sharpe's CAPM (Capital Asset Pricing Model) [42]:

$$\bar{R} = \alpha + \beta \bar{R}_m \tag{2}$$

where $\bar{R}$ is the expected return of the stock, $\bar{R}_m$ is the return of a market index, $\alpha$ is a variable representing the component of the stock's return that is independent of the market's performance, and $\beta$ is a constant that measures the expected change in $\bar{R}$ given a change in the market return $\bar{R}_m$. Although CAPM (2) provides a simple and insightful model for expected return, it considers only the market return $\bar{R}_m$ and ignores other important factors that may influence the expected return. Therefore, the so-called *factor models* [16]-[18] were proposed in the literature:

$$\bar{R} = \alpha + \sum_{j=1}^{n} \beta_j \bar{F}_j \tag{3}$$

where $\alpha$, $\beta_j$ are constants, and $\bar{F}_j$ are factors such as market return $\bar{R}_m$, earning-price ratios, dividend-price ratios, interest rates, stock momentum, corporate payout, etc., that may influence the stock return. Although the factor model (3) has been extensively studied in the mainstream financial economic literature for more than forty years, it was concluded in an influential study [23] that "… *these models have predicted poorly both in-sample and out-of-sample for 30 years now, …, the profession has yet to find some variable that has meaningful and robust empirical equity premium forecasting power*."

Humans are social animals (as Aristotle famously noticed), and our opinions about something are strongly influenced by the opinions of our peers. This is particularly true in the investment world. For example, in his bestseller Irrational Exuberance [44], Shiller wrote:

> "A fundamental observation about human society is that people who communicate regularly with one another think similarly. There is at any place and in any time a Zeitgeist, a spirit of the times. … Word-of-mouth transmission of ideas appears to be an important contributor to the day-to-day or hour-to-hour stock market fluctuations."

There have been many empirical studies that documented the evidence that the expectations of investors are strongly influenced by the opinions of other investors in their social networks, and a small samples of these studies are as follows: Shiller and Pound [46] surveyed 131 individual investors and asked them about what had drawn their attention to the stock they had most recently purchased, many of these investors named a personal contact such as a friend or a relative; Hong, Kubik and Stein [28] studied word-of-mouth effects among mutual fund managers and found that "A manager is more likely to hold (or buy, or sell) a particular stock in any quarter if other managers in the same city are holding (or buying, or selling) that same stock;" Feng and Seasholes [19] presented evidence of herding effects among individual investors who hold individual brokerage accounts in China; Coval and Moskowitz [9], [10] studied manager-level data and found that fund managers tend to overweight nearby companies, suggesting a link between geographic proximity and information transmission; Cohen, Frazzini and Malloy [11] found that portfolio managers place larger bets on firms they are connected to through shared education networks; and more recently, Pool, Stoffman and Yonker [41] discovered that fund managers with similar ethnic background have more similar holdings and trades, and Frydman [20] showed, using neural data collected from an experimental asset market, that "the subjects with the strongest neural sensitivity to a peer's change in wealth exhibit the largest peer effects in their trading behavior."

The goal of this paper is to develop a mathematical model for the word-of-mouth effects through social networks and explore how the changes of investors' social networks influence the stock price dynamics. First, we use a Gaussian fuzzy set to model the stock price expectation of an investor, where the center of the Gaussian fuzzy set is the expected price and the standard deviation of the Gaussian fuzzy set represents the uncertainty about the expected price. Then, based on a similarity measure between Gaussian fuzzy sets, we propose a *bounded confidence fuzzy opinion network* (BCFON) to model the social connection of the investors, where only those investors whose fuzzy expectations are close to each other are connected, and the investors in a connected group update their fuzzy expectations as the averages of the previous fuzzy expectations of their neighbors. Finally, the fuzzy expectations from the BCFON are used as inputs to drive the stock price dynamics.

The price dynamic models of this paper belong to the class of *agent-based models* (ABMs) for finance and economy [7], [26] where the agents are investors communicating through social networks [32], [53]. There are two main approaches to studying the dynamic changes of stock prices: the random walk model [3] and the agent-based models [49]. The random walk model assumes that the returns (relative changes of prices) are random and driven by an i.i.d. random process. The advantage of the random walk model is that it is a good first-approximation to real stock prices and provides a simple model based on which other important problems in finance and economy can be studied in a mathematically rigorous fashion (e.g. [30]). In fact the whole field of mathematical finance [47] is built mostly on the random walk model (whenever the dynamic changes of stock prices are concerned). A problem of the random walk model is that it does not reveal the mechanism of how the prices change. Furthermore, some key phenomena robustly observed in real stock prices, such as trends, cannot be explained by the random walk model. The agent-based models try to solve these problems by following a bottom-up approach to model directly the operations of different types of traders such as fundamentalists and chartists [26]. There is a large literature of agent-based models for stock prices [49], ranging from simple heuristic models [35] to sophisticated switching among different types of traders [38]. Recently, technical trading rules were converted into agent-based models through fuzzy system techniques to model the dynamic changes of stock prices [51]. The novelty of the price dynamic models of this paper is that they provide the mathematical details of how communications among investors through social networks influence the stock price dynamics and vice versa.

This paper is organized as follows. In Section II, the basics of the price dynamic model, including the basic price dynamic equation, the fuzzy set formulation of the expected prices and their uncertainties, and the structure of the Bounded Confidence Fuzzy Opinion Network (BCFON), are proposed. In Section III, the detailed evolution formulas of the BCFON are derived and the convergence properties of the BCFON are studied. Section IV summarizes the dynamic equations of the price dynamic model and extends it to model the actions of three typical traders: followers, contrarians and manipulators. In Section V, Monte Carlo simulations are performed to illustrate the price series generated by the price dynamic models and to study how the parameters of the models influence the key properties of the price dynamic models. In Section VI, the basic price dynamic model is modified by introducing the combined expected price and the combined uncertainty of all the investors, and the modified price dynamic model is identified based on the real daily closing prices of fifteen top banking and real estate stocks listed in the Hong Kong Stock Exchange for the recent two years from Dec. 5, 2013 to Dec. 4, 2015. Finally, Section VII concludes the paper, and the Appendix contains the proofs of all the theorems in this paper.

## II. The Basics of Price Dynamic Model

Suppose there are $n$ investors trading a stock at time $t$. Let $p_t$ be the price of the stock at time $t$ and $ed_{i,t}$ be the *excess demand* of investor $i$ at time $t$, then the price dynamics can be modeled as follows:

$$\ln(p_{t+1}) = \ln(p_t) + \sum_{i=1}^{n} ed_{i,t} + \varepsilon_t \qquad (4)$$

where $\ln(p_{t+1}) - \ln(p_t) \approx \frac{p_{t+1}-p_t}{p_t}$ is the relative change (return) of the price, $\varepsilon_t$ is an i.i.d. zero-mean Gaussian noise, and the excess demand $ed_{i,t}$ is chosen as:

$$ed_{i,t} = \frac{a_i\big[\ln(\bar{p}_{i,t}) - \ln(p_t)\big]}{\sigma_{i,t}} \qquad (5)$$

where $\bar{p}_{i,t}$ is investor $i$'s expectation of the stock price, $\sigma_{i,t}$ characterizes the uncertainty about the expectation $\bar{p}_{i,t}$, and $a_i$ is a positive constant characterizing the strength of investor $i$. The meaning of the price dynamic model (4) and (5) is that if the price $p_t$ is lower (higher) than the expected price $\bar{p}_{i,t}$, meaning that the stock is undervalued (overvalued) according to investor $i$'s judgment, then investor $i$ will buy (sell) the stock, leading to a positive (negative) excess demand $ed_{i,t}$ for the stock with its magnitude propositional to the relative difference $\ln(\bar{p}_{i,t}) - \ln(p_t) \approx \frac{\bar{p}_{i,t}-p_t}{p_t}$ between the expected price $\bar{p}_{i,t}$ and the real price $p_t$ and inversely propositional to the uncertainty $\sigma_{i,t}$ about the expected price. When the strength parameters $a_i$ in (5) are all equal to zero, the price dynamic equation (4) is reduced to the random walk model $\ln(p_{t+1}) = \ln(p_t) + \varepsilon_t$ (a combined quantity with $a_i$ will be estimated based on real stock price data in Section VI).

Since the expected prices $\bar{p}_{i,t}$ are human opinions about the future stock prices which are inherently uncertain, their uncertainties $\sigma_{i,t}$ should be considered simultaneously with the $\bar{p}_{i,t}$. Therefore, we use a single fuzzy set $P_{i,t}$ with Gaussian membership function:

$$\mu_{P_{i,t}}(x) = e^{-\frac{|x-\bar{p}_{i,t}|^2}{\sigma_{i,t}^2}} \qquad (6)$$

to model the expected price of the stock [56], where the center $\bar{p}_{i,t}$ of the Gaussian fuzzy set is the expected price itself and the standard deviation $\sigma_{i,t}$ of the Gaussian fuzzy set represents the uncertainty about the expected price. With the fuzzy set formulation (6), the price expectation becomes a fuzzy number $P_{i,t}$ over the domain of the price (we call it a *fuzzy expectation*), and our task is to determine a meaningful mathematical model for the evolution of the fuzzy expectation $P_{i,t}$.

As discussed in the Introduction, an investor's opinion about the price of a stock is strongly influenced by the opinions of their neighbors in some kind of social networks. In this paper, we will develop a *Bounded Confidence Fuzzy Opinion Network* (BCFON) to model the evolution of the fuzzy price expectation $P_{i,t}$, where the basic argument is the psychological discovery that people are mostly influenced by those with similar opinions [24]. We now start with some definitions and then propose the details of the BCFON.

**Definition 1:** A *fuzzy opinion* is a Gaussian fuzzy set $X$ with membership function $\mu_X(x) = e^{-\frac{|x-c|^2}{\sigma^2}}$ where the center $c \in R$ represents the opinion and the standard deviation $\sigma \in R_+$ characterizes the uncertainty about the opinion $c$. In particular, the fuzzy expectation $P_{i,t}$ with Gaussian membership function $\mu_{P_{i,t}}(x) = e^{-\frac{|x-\bar{p}_{i,t}|^2}{\sigma_{i,t}^2}}$ is a fuzzy opinion. ∎

**Definition 2:** The *closeness between fuzzy opinions $X_1$ and $X_2$* (with membership functions $\mu_{X_1}(x) = e^{-\frac{|x-c_1|^2}{\sigma_1^2}}$ and $\mu_{X_2}(x) = e^{-\frac{|x-c_2|^2}{\sigma_2^2}}$), denoted as $\max(X_1 \cap X_2)$, is defined as the height of the intersection of the two fuzzy sets $X_1$ and $X_2$, i.e.,

$$\max(X_1 \cap X_2) = \max_x \min[\mu_{X_1}(x), \mu_{X_2}(x)] = e^{-\frac{|c_1-c_2|^2}{(\sigma_1+\sigma_2)^2}} \qquad (7)$$

as illustrated in Fig. 1. ∎

**Definition 3:** A *Fuzzy Opinion Network* (FON[1]) is a connection of a number of Gaussian nodes, possibly through weighted-sum, delay and logic operation elements, where a *Gaussian node* is a 2-input-1-output fuzzy opinion $X_i$ with Gaussian membership function $\mu_{X_i}(x) = e^{-\frac{|x-C_i|^2}{\Sigma_i^2}}$ whose center $C_i$ and standard deviation $\Sigma_i$ are two input fuzzy sets to the node and the fuzzy set $X_i$ itself is the output of the node. A node is also called an *agent* or an *investor* throughout this paper. ∎

---

[1] The concept of FON was first introduced in [52] where the basic FON theory was developed through a detailed study of 13 representative FONs ranging from the basic static connections to the more general feedback, time-varying and state-dependent connections.

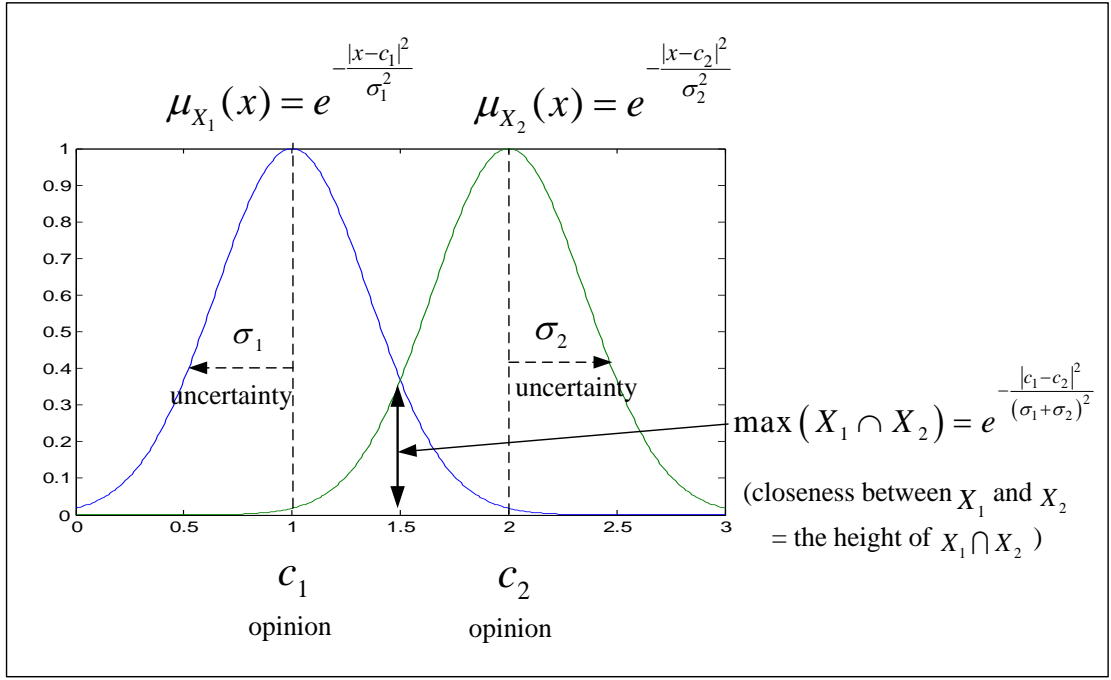

Fig. 1: Closeness between fuzzy opinions $X_1$ and $X_2$: $\max(X_1 \cap X_2)$.

**Definition 4:** Let $X, V$ be fuzzy sets defined on the universes of discourse $\Omega_X$, $\Omega_V$, respectively. A *conditional fuzzy set*, denoted as $X|V$, is a fuzzy set defined on $\Omega_X$ with the membership function:

$$\mu_{X|V}(x|V), \quad x \in \Omega_X \qquad (8)$$

depending on the fuzzy set $V$ whose membership function is $\mu_V(v)$ with $v \in \Omega_V$. Given $\mu_{X|V}(x|V)$ and $\mu_V(v)$, the membership function of the *unconditional fuzzy set $X$* is obtained from Zadeh's Compositional Rule of Inference [54] as follows:

$$\mu_X(x) = \max_{v \in \Omega_V} min[\mu_{X|V}(x|v), \mu_V(v)] \qquad (9)$$

∎

With these definitions, we are now ready to construct the bounded confidence fuzzy opinion networks as follows.

**Construction of Bounded Confidence Fuzzy Opinion Network[2] (BCFON):** A *bounded confidence fuzzy opinion network* is a dynamic connection of *n* fuzzy nodes $X_i(t)$ ($i = 1,2, \ldots, n$) with membership functions $\mu_{X_i(t)}(x) = e^{-\frac{|x-C_i(t)|^2}{u_i^2(t)}}$, where the center input $C_i(t+1)$ and the standard deviation input $u_i(t+1)$ to node $i$ at time $t+1$ ($t = 0,1,2,\ldots$) are determined as follows: the center input $C_i(t+1)$ is a weighted average of the outputs $X_j(t)$ of the $n$ fuzzy nodes at the previous time point $t$:

$$C_i(t+1) = \sum_{j=1}^{n} w_{ij}(t) X_j(t) \tag{10}$$

with the weights

$$w_{ij}(t) = \begin{cases} \frac{1}{|N_i(t)|}, & j \in N_i(t) \\ 0, & j \notin N_i(t) \end{cases} \tag{11}$$

where $N_i(t)$ ($i = 1, \ldots, n$) is the collection of nodes that are connected to node $i$ at time $t$, defined as:

$$N_i(t) = \left\{ j \in \{1, \ldots, n\} \mid \max\left(X_i(t) \cap X_j(t)\right) \geq d_i \right\} \tag{12}$$

where $d_i \in [0,1]$ are constants and $|N_i(t)|$ denotes the number of elements in $N_i(t)$; and, the standard deviation input $u_i(t+1)$ are determined according to one of the three schemes:

(a) *Local Reference*:

$$u_i(t+1) = b \left| \text{Cen}(X_i(t)) - \frac{1}{|N_i(t)|} \sum_{j \in N_i(t)} \left( \text{Cen}(X_j(t)) \right) \right| \tag{13}$$

---

[2] The BCFON proposed in [52] is different from the BCFON constructed here: the weights in the BCFON of [52] are weighted averages of the distances between the fuzzy opinions, while the weights here are equal within a group as defined in (11); also, the External Reference scheme (15) for the uncertainty input $u_i$ is new.

(b) *Global Reference*:

$$u_i(t+1) = b \left| \text{Cen}(X_i(t)) - \frac{1}{n} \sum_{j=1}^{n} \text{Cen}(X_j(t)) \right| \tag{14}$$

(c) *External Reference*:

$$u_i(t+1) = b \left| \text{Cen}(X_i(t)) - g(t) \right| \tag{15}$$

where $\text{Cen}(X_i)$ denotes the center of fuzzy set $X_i$, $g(t)$ is an external signal (e.g., it may be the stock price $p_t$ from the price dynamic equation (4)) and $b$ is a positive scaling constant. The initial fuzzy opinions $X_i(0)$ ($i = 1,2, \ldots, n$) are Gaussian fuzzy sets $\mu_{X_i(0)}(x) = e^{-\frac{|x-x_{i0}|^2}{\sigma_{i0}^2}}$, where the initial opinions $x_{i0} \in R$ and the initial uncertainties $\sigma_{i0} \in R_+$ are given. ∎

Fig. 2 shows the BCFON connected to the price dynamic model (4), (5). The meaning of the state-dependent weights $w_{ij}(t)$ of (11) is that the weight $w_{ij}(t)$ from node $j$ to node $i$ is non-zero only when the two fuzzy opinions $X_i(t)$ and $X_j(t)$ are close enough to each other such that their closeness $\max\left(X_i(t) \cap X_j(t)\right) \geq d_i$, and each agent $i$ (node $i$) gives the same weight $\frac{1}{|N_i(t)|}$ for the $|N_i(t)|$ agents $N_i(t)$ that are connected to agent $i$. The meaning of the Local Reference scheme (13) for the standard deviation input $u_i(t+1)$ is that agent $i$ considers only the opinions of his neighbors in $N_i(t)$ and views the average of his neighbors' latest opinions $\frac{1}{|N_i(t)|}\sum_{j \in N_i(t)}\left(\text{Cen}\left(X_j(t)\right)\right)$ as the correct answer, therefore the closer his last opinion $\text{Cen}(X_i(t))$ is to the average of the latest opinions of his neighbors, the more confidence he has; this gives the choice of the standard deviation input in (13). The meaning of the Global Reference scheme (14) is that agent $i$ considers the opinions of all the agents and views the average of all the agents' opinions $\frac{1}{n}\sum_{j=1}^{n}\text{Cen}\left(X_j(t)\right)$ as the correct answer; the closer his opinion is to this correct answer, the less uncertain he is, and this leads to the choice of $u_i(t+1)$ in (14). The Global Reference case corresponds to the situation where a central agent collects the opinions from all the people involved and announces the averaged opinion back to all the people, such as the scenario in Keynes' Beauty Contest Theory (chapter 12 of [34]). The Local Reference case, on the other hand, refers to the scenario of decentralized control where a central agent does not exist and people know only the opinions of their neighbors. The External Reference (15) is a general one, where the reference signal $g(t)$ can be anything, such as $g(t) = p_t$ where $p_t$ is the price series generated by the price dynamic model (4). Our task next is to explore how the opinions $X_i(t)$ evolve in a BCFON.

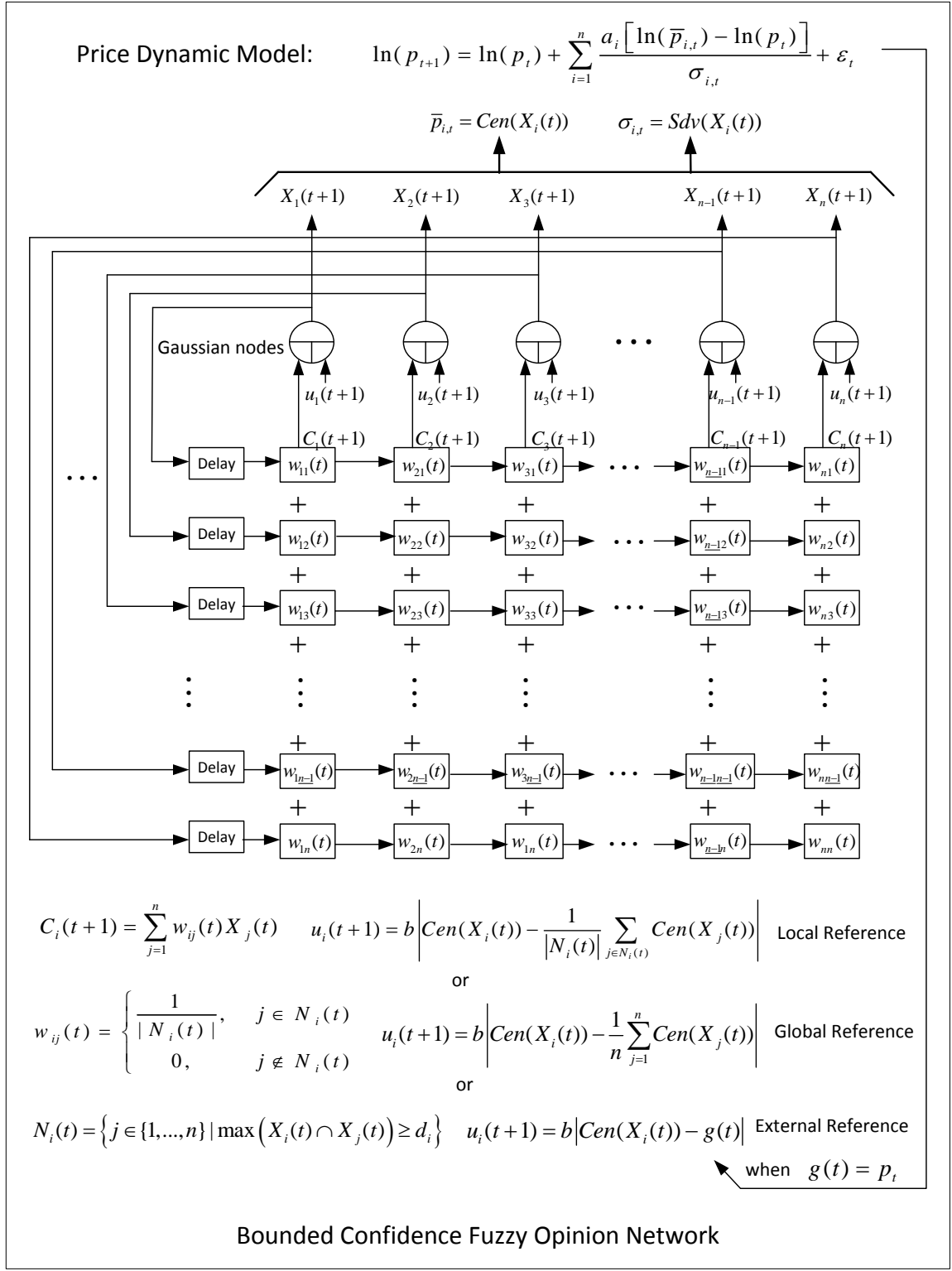

Fig. 2: The price dynamic model and the bounded confidence fuzzy opinion network (BCFON), where $\text{Cen}(X_i(t))$ and $\text{Sdv}(X_i(t))$ denote the center and the standard deviation of the Gaussian fuzzy set $X_i(t)$, respectively.

## III. The Evolution of BCFON and Convergence Results

Given the initial fuzzy opinions $X_i(0)$ ($i = 1, 2, \dots, n$) with $\mu_{X_i(0)}(x) = e^{-\frac{|x-x_{i0}|^2}{\sigma_{i0}^2}}$, the center input to node $i$ at the first time point $t = 1$ is

$$C_i(1) = \sum_{j=1}^{n} w_{ij}(0) X_j(0) \tag{16}$$

where the weights $w_{ij}(0)$, computed according to (11), are non-negative numbers and $\sum_{j=1}^{n} w_{ij}(0) = \sum_{j \in N_i(t)} \frac{1}{|N_i(t)|} = 1$. Now the question is: how to determine the fuzzy set $C_i(1)$ – a weighted average of $n$ Gaussian fuzzy sets. The following lemma from [52] gives the answer.

**Lemma 1:** Let $X_i$ ($i = 1, 2, \dots, n$) be fuzzy sets with Gaussian membership functions $\mu_{X_i}(x_i) = e^{-\frac{|x_i - c_i|^2}{\sigma_i^2}}$ and $w_i \geq 0$ be constant weights with $\sum_{i=1}^{n} w_i = 1$. Then,

$$Y_n = w_1 X_1 + w_2 X_2 + \cdots + w_n X_n \tag{17}$$

is a Gaussian fuzzy set with membership function:

$$\mu_{Y_n}(y_n) = e^{-\frac{|y_n - \sum_{i=1}^{n} w_i c_i|^2}{(\sum_{i=1}^{n} w_i \sigma_i)^2}} \tag{18}$$

∎

The proof of this lemma can be found in [52]. Applying Lemma 1 to (16), we have that the center input $C_i(1)$ to node $i$ at time point $t = 1$ is a Gaussian fuzzy set with membership function:

$$\mu_{C_i(1)}(x) = e^{-\frac{|x - \sum_{j=1}^{n} w_{ij}(1) x_{j0}|^2}{(\sum_{j=1}^{n} w_{ij}(1) \sigma_{j0})^2}} \tag{19}$$

To get the output $X_i(1)$ of node $i$ at time point $t = 1$, we need the following lemma.

**Lemma 2:** Let $X|V$ be a conditional Gaussian fuzzy set with membership function:

$$\mu_{X|V}(x|V) = e^{-\frac{|x - V|^2}{\sigma_X^2}} \tag{20}$$

where $\sigma_X$ is a positive constant and $V$ is a Gaussian fuzzy set with membership function:

$$\mu_V(v) = e^{-\frac{|v - c_V|^2}{\sigma_V^2}} \tag{21}$$

where $c_V$ and $\sigma_V$ are positive constants. Then, the unconditional fuzzy set $X$ is Gaussian with membership function:

$$\mu_X(x) = e^{-\frac{|x-c_V|^2}{(\sigma_X+\sigma_V)^2}} \tag{22}$$

∎

The proof of Lemma 2 is given in the Appendix. Since the uncertainty input $u_i(1)$ to node $i$ at time point $t = 1$ is a non-fuzzy real number according to (13), (14) or (15)

$$\left( u_i(1) = \left|x_{i0} - \frac{1}{|N_i(0)|}\sum_{j \in N_i(0)}(x_{j0})\right| \right., \quad N_i(0) = \left\{ j \in \{1, \ldots, n\} \,|\, e^{-\frac{|x_{i0}-x_{j0}|^2}{(\sigma_{i0}+\sigma_{j0})^2}} \geq d_i \right\},$$

$u_i(1) = \left|x_{i0} - \frac{1}{n}\sum_{j=1}^{n} x_{j0}\right|$, or $u_i(1) = |x_{i0} - g(0)|$ ), applying Lemma 2 to

$\mu_{X_i(1)|C_i(1)}(x) = e^{-\frac{|x-C_i(1)|^2}{(u_i(1))^2}}$ and (19) gives

$$\mu_{X_i(1)}(x) = e^{-\frac{\left|x-\sum_{j=1}^{n} w_{ij}(0)x_{j0}\right|^2}{\left(\sum_{j=1}^{n} w_{ij}(0)\sigma_{j0}+u_i(1)\right)^2}} \tag{23}$$

Since $X_i(1)$ ($i = 1, \ldots, n$) are still Gaussian, we can continue the above process for $t = 2, 3, \ldots$ to get the general results of the BCFON evolution, which is summarized in the following theorem.

**Theorem 1:** The fuzzy opinions $X_i(t+1)$ ($i = 1, \ldots, n; t = 0,1,2,\ldots$) in the BCFON of Fig. 2 are Gaussian fuzzy sets:

$$\mu_{X_i(t+1)}(x) = e^{-\frac{|x-\text{Cen}(X_i(t+1))|^2}{(\text{Sdv}(X_i(t+1)))^2}} \tag{24}$$

where the opinions $\text{Cen}(X_i(t+1)) \in R$ and their uncertainties $\text{Sdv}(X_i(t+1)) \in R_+$ are evolving according to the following dynamic equations:

$$\text{Cen}(X_i(t+1)) = \sum_{j=1}^{n} w_{ij}(t)\text{Cen}(X_j(t)) \tag{25}$$

$$\text{Sdv}(X_i(t+1)) = \sum_{j=1}^{n} w_{ij}(t)\text{Sdv}(X_j(t)) + u_i(t+1) \tag{26}$$

where the weights

$$w_{ij}(t) = \begin{cases} \frac{1}{|N_i(t)|}, & j \in N_i(t) \\ 0, & j \notin N_i(t) \end{cases} \tag{27}$$

$$N_i(t) = \left\{ j \in \{1, \ldots, n\} \mid e^{-\frac{|\text{Cen}(X_i(t)) - \text{Cen}(X_j(t))|^2}{(\text{Sdv}(X_i(t)) + \text{Sdv}(X_j(t)))^2}} \geq d_i \right\} \tag{28}$$

and the uncertainty input

$$u_i(t+1) = b \left| \text{Cen}(X_i(t)) - \frac{1}{|N_i(t)|} \sum_{j \in N_i(t)} \left( \text{Cen}(X_j(t)) \right) \right| \tag{29}$$

for *Local Reference*,

$$u_i(t+1) = b \left| \text{Cen}(X_i(t)) - \frac{1}{n} \sum_{j=1}^{n} \text{Cen}(X_j(t)) \right| \tag{30}$$

for *Global Reference*, or

$$u_i(t+1) = b \left| \text{Cen}(X_i(t)) - g(t) \right| \tag{31}$$

for *External Reference* with initial condition $\text{Cen}(X_i(0)) = x_{i0}$ (initial opinion of agent $i$) and $\text{Sdv}(X_i(0)) = \sigma_{i0}$ (uncertainty about the initial opinion), where $0 \leq d_i \leq 1$, $b > 0$ are constants. ∎

The proof of Theorem 1 is given in the Appendix. The following theorem gives the convergence results of BCFON (25)-(28) with Local Reference (29).

**Theorem 2:** Consider the BCFON dynamics (25)-(28) with Local Reference (29) and assume that the $d_i$'s in (28) of all the investors are the same, i.e., $d_i = d \in (0,1)$ for all $i = 1, \ldots, n$. Given initial condition $\text{Cen}(X_i(0)) = x_{i0} \in R$ and $\text{Sdv}(X_i(0)) = \sigma_{i0} \in R_+$, the $n$ investors converge to $q$ disjoint groups $I_1, \ldots, I_q$ ( $I_1 \cup \cdots \cup I_q = \{1, \ldots, n\}$ and $I_i \cap I_j = \emptyset$ for $i \neq j$) with the opinions $\text{Cen}(X_i(t))$ and their uncertainties $\text{Sdv}(X_i(t))$ of the investors in the same group $G_k$ converging to the same values $x_{k\infty}$ and $\sigma_{k\infty}$, respectively, as $t$ goes to infinity, i.e.,

$$\lim_{t \to \infty} \text{Cen}(X_i(t)) = x_{k\infty} \tag{32}$$

$$\lim_{t \to \infty} \text{Sdv}(X_i(t)) = \sigma_{k\infty} \tag{33}$$

for $i \in I_k$, $k = 1, \ldots, q$.[3] Furthermore, the convergence of (32) and (33) are achieved in finite steps, i.e., there exists $t_N$ such that $\text{Cen}(X_i(t)) = x_{k\infty}$ and $\text{Sdv}(X_i(t+1)) = \sigma_{k\infty}$, $i \in I_k$, $k = 1, \ldots, q$, for all $t > t_N$. ∎

The proof of Theorem 2 is given in the Appendix. The next theorem gives the convergence results of BCFON (25)-(28) with Global Reference (30).

**Theorem 3:** Consider the BCFON dynamics (25)-(28) with Global Reference (30) and assume that the $d_i$'s in (28) of all the investors are the same, i.e., $d_i = d \in (0,1)$ for all $i = 1, \ldots, n$. Given initial condition $\text{Cen}(X_i(0)) = x_{i0} \in R$ and $\text{Sdv}(X_i(0)) = \sigma_{i0} \in R_+$, the $n$ investors converge to a consensus, i.e.,

$$\lim_{t \to \infty} \text{Cen}(X_i(t)) = x_\infty \tag{34}$$

$$\lim_{t \to \infty} \text{Sdv}(X_i(t)) = \sigma_\infty \tag{35}$$

for all $i = 1, 2, \ldots, n$, where $x_\infty$, $\sigma_\infty$ are constants. Furthermore, the convergence of (34) and (35) is achieved in finite steps, i.e., there exists $t_N$ such that $\text{Cen}(X_i(t)) = x_\infty$ and $\text{Sdv}(X_i(t+1)) = \sigma_\infty$ for all $i = 1, \ldots, n$ and all $t > t_N$. ∎

The proof of Theorem 3 is given in the Appendix. The next theorem gives the convergence results of BCFON (25)-(28) with External Reference (31).

**Theorem 4:** Consider the BCFON dynamics (25)-(28) with External Reference (31) and assume that the $d_i$'s in (28) of all the investors are the same, i.e., $d_i = d \in (0,1)$ for all $i = 1, \ldots, n$. Given initial condition $\text{Cen}(X_i(0)) = x_{i0} \in R$ and $\text{Sdv}(X_i(0)) = \sigma_{i0} \in R_+$, the opinions $\text{Cen}(X_i(t))$ of the $n$ investors converge to the same value $x_\infty$, i.e.,

$$\lim_{t \to \infty} \text{Cen}(X_i(t)) = x_\infty \tag{36}$$

for all $i = 1, 2, \ldots, n$, and the convergence of (36) is achieved in finite steps, i.e., there exists $t_N$ such that $\text{Cen}(X_i(t)) = x_\infty$ for all $i = 1, \ldots, n$ and all $t > t_N$. For the uncertainties $\text{Sdv}(X_i(t))$, we have $\text{Sdv}(X_i(t+1)) = \sigma_{t+1}$ for all $i = 1, \ldots, n$ after $t > t_N$, and

$$\sigma_{t+1} = \sigma_t + b|x_\infty - g(t)| \tag{37}$$

from which we get

$$\lim_{t\to\infty} \text{Sdv}(X_i(t)) = \sigma_{t_N+1} + \sum_{t=t_N+1}^{\infty} b|x_\infty - g(t)| \tag{38}$$

which may be a finite number or go to infinity depending on the values of the external signal $g(t)$ and the opinion consensus $x_\infty$. ∎

---

[3] The number of converged groups $q$ and the specific members of the converged groups $G_k$ depend on the threshold $d$ and the initial distributions of $\text{Cen}(X_i(0))$ and $\text{Sdv}(X_i(0))$ in a very complex fashion. The simulations in Section V will show some examples of $q$ and $G_k$ in certain typical situations, but the general mathematical results for $q$ and $G_k$ are currently not available.

The proof of Theorem 4 is given in the Appendix. In the next section, we apply Theorem 1 to the basic price dynamic model to obtain the final price dynamic models.

## IV. The Final Price Dynamic Models

Applying Theorem 1 to the basic price dynamic model (4) and (5) with $\text{Cen}(X_i(t)) = \bar{p}_{i,t}$ and $\text{Sdv}(X_i(t)) = \sigma_{i,t}$, we obtain the final price dynamic model as follows:[4]

**The Stock Price Dynamic Model:** Consider $n$ investors trading a stock whose price at time $t$ is $p_t$ ($t = 0,1,2,...$). Suppose the $n$ investors are connected through the Bounded Confidence Fuzzy Opinion Network in Fig. 2 whose dynamics are summarized in Theorem 1. Then, with initial condition $\bar{p}_{i,0}$ (initial expected price of the stock from investor $i$), $\sigma_{i,0}$ (uncertainty of investor $i$ about the initial expected price $\bar{p}_{i,0}$) and $p_0$ (stock price at time 0) for $i = 1,2,...,n$, the stock price $p_t$ ($t = 0,1,2,...$) is changing according to the following dynamic equations:

$$\ln(p_{t+1}) = \ln(p_t) + \sum_{i=1}^{n} \frac{a_i[\ln(\bar{p}_{i,t}) - \ln(p_t)]}{\sigma_{i,t}} + \varepsilon_t \tag{39}$$

$$\bar{p}_{i,t+1} = \frac{1}{|N_i(t)|} \sum_{j\in N_i(t)} \bar{p}_{j,t} \tag{40}$$

$$\sigma_{i,t+1} = \frac{1}{|N_i(t)|} \sum_{j\in N_i(t)} \sigma_{j,t} + u_i(t+1) \tag{41}$$

where $a_i > 0$ are constants, $\varepsilon_t$ is a zero-mean i.i.d. Gaussian random process with standard deviation $\sigma_\varepsilon$,

---

[4] Specifically, substituting (5) into (4) we obtain (39); substituting $\text{Cen}(X_i(t)) = \bar{p}_{i,t}$ and $\text{Sdv}(X_i(t)) = \sigma_{i,t}$ into (25) and (26) with weights $w_{ij}(t)$ given by (27), we obtain (40) and (41); and, (42)-(45) are just (28)-(31) with $\text{Cen}(X_i(t)) = \bar{p}_{i,t}$ and $\text{Sdv}(X_i(t)) = \sigma_{i,t}$.

$$N_i(t) = \left\{ j \in \{1, \ldots, n\} \mid e^{-\frac{|\bar{p}_{i,t} - \bar{p}_{j,t}|^2}{(\sigma_{i,t} + \sigma_{j,t})^2}} \geq d_i \right\} \tag{42}$$

is the neighbors of investor $i$ at time $t$, $0 \leq d_i \leq 1$, $|N_i(t)|$ denotes the number of elements in $N_i(t)$, and the uncertainty input $u_i(t+1)$ to investor $i$ is chosen according to one of the following three schemes:

(a) *Local Reference*:

$$u_i(t+1) = b \left| \bar{p}_{i,t} - \frac{1}{|N_i(t)|} \sum_{j \in N_i(t)} \bar{p}_{j,t} \right| \tag{43}$$

(b) *Global Reference*:

$$u_i(t+1) = b \left| \bar{p}_{i,t} - \frac{1}{n} \sum_{j=1}^{n} \bar{p}_{j,t} \right| \tag{44}$$

(c) *Real Price Reference*:

$$u_i(t+1) = b |\bar{p}_{i,t} - p_t| \tag{45}$$

where $b > 0$ is a scaling constant. Note that the Real Price Reference (45) is the External Reference (31) with the external signal $g(t)$ chosen as the real price $p_t$. ∎

The following theorem gives the convergence properties of the price dynamic model (39)-(45).

**Theorem 5:** Consider the stock price dynamic model (39)-(45). According to Theorems 2, 3 and 4, let $\bar{p}_{i,\infty}$, $\sigma_{i,\infty}$ be the converged values of $\bar{p}_{i,t}$, $\sigma_{i,t}$ with Local Reference, $\bar{p}_\infty$, $\sigma_\infty$ be the consensus reached by $\bar{p}_{i,t}$, $\sigma_{i,t}$ with Global Reference, and $\bar{p}_\infty$ be the converged expected price with Real Price Reference.

(a) For Local Reference (43), if $0 < \sum_{i=1}^{n} \left( \frac{a_i}{\sigma_{i,\infty}} \right) < 2$, then the expected value of the log price $E\{\ln(p_t)\}$ converges to a constant as $t$ goes to infinity, specifically:

$$\lim_{t \to \infty} E\{\ln(p_t)\} = \frac{\sum_{i=1}^{n} \left( \frac{a_i \ln(\bar{p}_{i,\infty})}{\sigma_{i,\infty}} \right)}{\sum_{i=1}^{n} \left( \frac{a_i}{\sigma_{i,\infty}} \right)} \tag{46}$$

(b) For Global Reference (44), if $0 < \frac{1}{\sigma_\infty}\sum_{i=1}^{n} a_i < 2$, then the expected value of the log price $E\{\ln(p_t)\}$ converges to $\ln(\bar{p}_\infty)$, i.e.,

$$\lim_{t\to\infty} E\{\ln(p_t)\} = \ln(\bar{p}_\infty) \tag{47}$$

(c) For Real Price Reference (45), the expected value of the log price $E\{\ln(p_t)\}$ always converges, but in general not to $\ln(\bar{p}_\infty)$, i.e.,

$$\lim_{t\to\infty} E\{\ln(p_t)\} = \ln(c) \tag{48}$$

where $c$ is a constant and is in general not equal to the consensus $\bar{p}_\infty$ reached by the $n$ investors. ∎

The proof of Theorem 5 is given in the Appendix.

In the real financial world, investors are heterogeneous. Although different traders use different trading strategies, roughly speaking the majority of investors may be classified into three types: followers, contrarians, or manipulators. *Followers* begin to trade (buy or sell) when their opinions are close enough to the majority's opinion, *contrarians* begin to buy or sell when their opinions are against the majority's opinion, and *manipulators* do not change their opinions at all. Experienced traders should observe very frequently in their daily trading that it is the interplay among trend followers, contrarians and manipulators that drives the prices into chaos (see [51] for an in-depth analysis of this phenomenon). We can easily modify the Stock Price Dynamic Model (39)-(45) to model the actions of followers, contrarians and manipulators, as follows:

**The Stock Price Dynamic Model with Followers:** The same as the Stock Price Dynamic Model (39)-(45) except that (39) is replaced by

$$\ln(p_{t+1}) = \ln(p_t) + \sum_{i=1}^{n} \frac{I_{t,i} a_i [\ln(\bar{p}_{i,t}) - \ln(p_t)]}{\sigma_{i,t}} + \varepsilon_t \tag{49}$$

where the indicator function $I_{t,i}$ is defined as

$$I_{t,i} = \begin{cases} 1, & \text{if } \left|\ln(\bar{p}_{i,t}) - \ln\left(\frac{1}{|N_i(t)|}\sum_{j\in N_i(t)} \bar{p}_{j,t}\right)\right| < c_i \\ 0, & \text{otherwise} \end{cases} \tag{50}$$

for Local Reference (43),

$$I_{t,i} = \begin{cases} 1 & \text{if } \left|\ln(\bar{p}_{i,t}) - \ln\left(\frac{1}{n}\sum_{j=1}^{n} \bar{p}_{j,t}\right)\right| < c_i \\ 0 & \text{otherwise} \end{cases} \quad (51)$$

for Global Reference (44), and

$$I_{t,i} = \begin{cases} 1 & \text{if } \left|\ln(\bar{p}_{i,t}) - \ln(p_t)\right| < c_i \\ 0 & \text{otherwise} \end{cases} \quad (52)$$

for Real Price Reference (45), where $c_i$ are positive constants. The meaning of the indicator functions (50)-(52) is that investor $i$ trades ($I_{t,i} = 1$) only when the relative difference between his expected price $\bar{p}_{i,t}$ and the majority's opinion $\frac{1}{|N_i(t)|}\sum_{j \in N_i(t)} \bar{p}_{j,t}$ for Local Reference, $\frac{1}{n}\sum_{j=1}^{n} \bar{p}_{j,t}$ for Global Reference or $p_t$ for Real Price Reference is less than $100 c_i \%$. ∎

**The Stock Price Dynamic Model with Contrarians:** The same as the Stock Price Dynamic Model with Followers except that the indicator function $I_{t,i}$ is chosen as

$$I_{t,i} = \begin{cases} 1, & \text{if } \left|\ln(\bar{p}_{i,t}) - \ln\left(\frac{1}{|N_i(t)|}\sum_{j \in N_i(t)} \bar{p}_{j,t}\right)\right| > c_i \\ 0, & \text{otherwise} \end{cases} \quad (53)$$

for Local Reference (43),

$$I_{t,i} = \begin{cases} 1 & \text{if } \left|\ln(\bar{p}_{i,t}) - \ln\left(\frac{1}{n}\sum_{j=1}^{n} \bar{p}_{j,t}\right)\right| > c_i \\ 0 & \text{otherwise} \end{cases} \quad (54)$$

for Global Reference (44), or

$$I_{t,i} = \begin{cases} 1 & \text{if } \left|\ln(\bar{p}_{i,t}) - \ln(p_t)\right| > c_i \\ 0 & \text{otherwise} \end{cases} \quad (55)$$

for Real Price Reference (45), where $c_i$ are positive constants. The meaning of the indicator functions (53)-(55) is that investor $i$ trades ($I_{t,i} = 1$) only when the relative difference between his expected price $\bar{p}_{i,t}$ and the majority's opinion $\frac{1}{|N_i(t)|}\sum_{j\in N_i(t)} \bar{p}_{j,t}$ for Local Reference, $\frac{1}{n}\sum_{j=1}^{n} \bar{p}_{j,t}$ for Global Reference or $p_t$ for Real Price Reference is greater than $100c_i\%$. ∎

**The Stock Price Dynamic Model with Manipulators:** The same as the Stock Price Dynamic Model (39)-(45) except that for some $I_0 \subset \{1, \ldots, n\}$ we have $d_i = 1$ for $i \in I_0$, i.e., the investors in $I_0$ do not change their expected prices at all; we call these investors *manipulators*. ∎

Can the manipulators successfully change the other investors' expected prices towards the manipulated values? What happens when there are more than one manipulators? The following theorem proves that in the Real Price Reference case the manipulators can indeed manipulate other investors' opinions, but when there are more than one manipulators with different manipulation goals, the other investors' expected prices will converge to the average of these manipulation targets; this shows that the biggest enemy of a manipulator is not the ordinary investors, but the other manipulators.

**Theorem 6:** Consider the Stock Price Dynamic Model (39)-(42) with Real Price Reference (45). Suppose there are $m$ investors $I_0 = \{1, \ldots, m\}$ who are manipulators, i.e., $d_i = 1$ for $i \in I_0$, and the remaining $n - m$ investors $I_1 = \{m+1, \ldots, n\}$ are ordinary investors with equal confidence bounds, i.e., $d_i = d < 1$ for $i \in I_1$. Assume that the initial expected prices of the $n$ investors $\bar{p}_{i,0}$ are different and a manipulator never takes other investor as neighbor, then:

(a) The expected prices of the manipulators never change, i.e.,

$$\bar{p}_{i,t} = \bar{p}_{i,0} \tag{56}$$

for $i \in I_0$ and $t = 1, 2, \ldots$, and the uncertainties about these expected prices change according to

$$\sigma_{i,t+1} = \sigma_{i,t} + b|\bar{p}_{i,0} - p_t| \tag{57}$$

where $i \in I_0$.

(b) For a typical realization of the price series $p_t$ such that $p_t$ does not converge to a constant as $t$ goes to infinity, the expected prices $\bar{p}_{j,t}$ of the ordinary investors $j \in I_1$ converge to the same $\bar{p}_t$ in finite time which converges to the average of the manipulators' expected prices, i.e., there exists $t_N$ such that $\bar{p}_{j,t} = \bar{p}_t$ for all $j \in I_1$ when $t > t_N$ and

$$\lim_{t\to\infty} \bar{p}_t = \frac{1}{m}\sum_{i=1}^{m} \bar{p}_{i,0} \tag{58}$$

The uncertainties $\sigma_{j,t}$ of the ordinary investors $j \in I_1$ converge to the same $\sigma_t$ in finite step which changes according to

$$\sigma_{t+1} = \frac{n-m}{n}\sigma_t + \frac{1}{n}\sum_{i=1}^{m}\sigma_{i,t} + b\left|\frac{1}{m}\sum_{i=1}^{m}\bar{p}_{i,0} - p_t\right| \tag{59}$$

after the finite step, i.e., there exists $t_N$ such that $\sigma_{j,t} = \sigma_t$ for all $j \in I_1$ when $t > t_N$, and the $\sigma_t$ follows (59) with $t > t_N + 1$. ∎

The proof of Theorem 6 is given in the Appendix.

We see from the Stock Price Dynamic Model (39)-(45) that the dynamics of the price $p_t$, the expected prices $\bar{p}_{i,t+1}$ and their uncertainties $\sigma_{i,t+1}$ are strongly nonlinear because the neighborhood sets $N_i(t)$ (42) and the uncertainty inputs $u_i(t+1)$ (43)-(45) are complex nonlinear functions of the past expected prices $\bar{p}_{j,t}$ and their uncertainties $\sigma_{j,t}$ of all the investors involved. The price dynamics with Followers, Contrarians or Manipulators are even more complex due to the added indicator operators $I_{t,i}$ (50)-(55) for the Followers and Contrarians cases and the non-homogenous choice of $d_i$ for the Manipulator case. Although we are fortunate to be able to prove some key convergence properties of these very complicated nonlinear dynamics in Theorems 2-6, it is helpful to see exactly what the price $p_t$, the expected prices $\bar{p}_{j,t}$ and the uncertainties $\sigma_{j,t}$ look like in typical runs of the price dynamic models to give us some concrete feelings about the dynamics of these time series. Furthermore, to see how the parameters of the models, such as $d_i$ (the thresholds to select neighbors), $n$ (the number of investors) and $b$ (the scaling parameter for the uncertainty inputs), influence the dynamics of the models, Monte Carlo simulations are one of the most effective methods. So in the next section we perform simulations of the price dynamic models developed in this section.

## V. Simulations

In this section, we perform simulations for the stock price dynamic models proposed in the last section. Specifically, Examples 1, 2 and 3 explore the Stock Price Dynamic Model (39)-(42) with Local Reference (43), Global Reference (44) and Real Price

Reference (45), respectively, and Examples 4, 5 and 6 simulate the Stock Price Dynamic Model with Followers, Contrarians and Manipulators, respectively.

**Example 1** (Local Reference): Consider the Stock Price Dynamic model (39)-(42) with Local Reference (43) and $n = 60$ investors. With parameters $a_i = 0.002$, $\sigma_\varepsilon = 0.02$, $d_i = 0.6$, $b = 1$ and initial $p_0 = 10$, $\bar{p}_{i,0}$ ($i = 1, \ldots, n$) uniformly distributed over the interval [5,25] ($\bar{p}_{i,0} = 5 + 20(\frac{i-1}{n-1})$, $i = 1, \ldots, n$) and their uncertainties $\sigma_{i,0}$ ($i = 1, \ldots, n$) drawn from a random uniform distribution over (0,1), Fig. 3 shows a simulation run of the dynamic model, where the top sub-figure plots the price $p_t$ (heavy line) and the pure random walk price (light line, obtained by setting $a_i = 0$ in the price equation (39)) for comparison, and the middle and bottom sub-figures plot the expected prices $\bar{p}_{i,t}$ ($i = 1, \ldots, n$) of the $n = 60$ investors and their uncertainties $\sigma_{i,t}$ ($i = 1, \ldots, n$), respectively. The converged mean price for the simulation run of Fig. 3, as computed from (46) of Theorem 5 (a), is $e^{\frac{\sum_{i=1}^{n}\left(\frac{a_i \ln(\bar{p}_{i,\infty})}{\sigma_{i,\infty}}\right)}{\sum_{i=1}^{n}\left(\frac{a_i}{\sigma_{i,\infty}}\right)}} = 13.88$. Comparing the price $p_t$ with the random walk price in the top sub-figure of Fig. 3 we see that the price $p_t$ is moving in a clear up-trend from the initial $p_0 = 10$ towards the final mean price around 13.88 when the investors exchange opinions and reach conclusions as shown in the middle and bottom sub-figures of Fig. 3, while the pure random walk price is just wandering around the initial price $p_0 = 10$ without any trend. As discussed in the Introduction that a main problem of the random walk model is that it cannot capture price trends that are commonly observed in real stock prices, here we see that our Stock Price Dynamic Model can produce trends with a clear explanation – a trend occurs during the process of exchanging opinions among investors and reaching new expected prices. Also, we see from Fig. 3 that with Local Reference (43), the 60 initially uniformly distributed expected prices eventually converge into four groups, and the investors in a group converge to the same expected price and the same uncertainty; furthermore, these convergences occur in finite steps, confirming the theoretical results of Theorem 2.

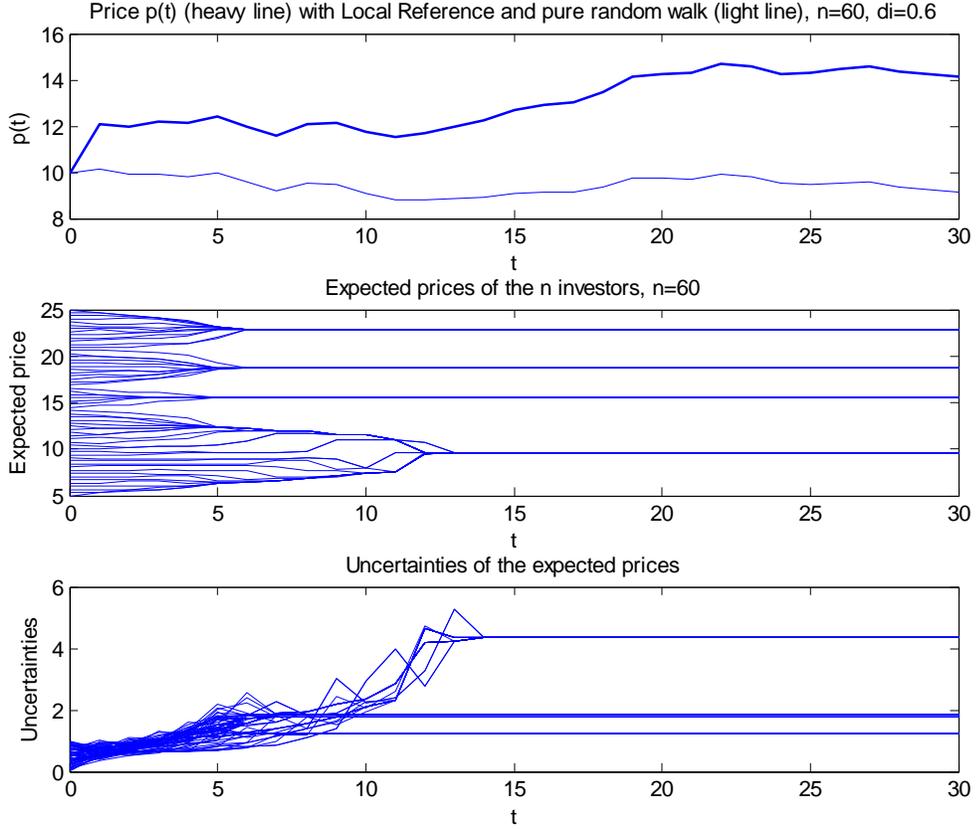

Fig. 3: A simulation run of the Stock Price Dynamic Model (39)-(42) with Local Reference (43) for $n = 60$ investors, where the top sub-figure plots the $p_t$ (heavy line) and pure random walk (light line), and the middle and bottom sub-figures plot the expected prices $\bar{p}_{i,t}$ and their uncertainties $\sigma_{i,t}$ ($i = 1, ..., n$), respectively.

To see the influence of the parameters $d_i$ (which are important parameters determining how the neighbors are formed) and $n$ (the number of investors which is also an important parameter influencing the characteristics of the system) on the number of converged groups (the $q$ in Theorem 2), we perform 100 Monte Carlo simulations and the results are summarized as follows: Fig. 4 shows the number of converged groups as function of $d_i$ for $n = 20, 40$ and $60$ investors (top, middle and bottom sub-figures of Fig. 4, respectively) in the 100 Monte Carlo simulation runs, Fig. 5 shows the number of converged groups as function of $n$ for $d_i = 0.75, 0.85$ and $0.95$ (top, middle and bottom sub-figures of Fig. 5, respectively) in the 100 Monte Carlo simulations, and Table 1 gives the average number of the converged groups and their standard deviations computed from 100 Monte Carlo simulations for different values of $d_i$ and $n$. We see from Figs. 4, 5 and Table 1 that the number of converged groups depends on $d_i$ and $n$ in a complicated nonlinear fashion. Specifically, from Fig. 4 we see that although it is generally true that the larger the $d_i$, the more the number of converged group, in some critical regions (which are different for different $n$) a small increase of $d_i$ results in a big increase of the number of converged groups. From Fig. 5 we see that as the number of investors $n$ increases, the number of converged groups first increases but then decreases after reaching some maximum numbers (which are different for different $d_i$); an explanation of this phenomenon is that when $n$ is very small, all investors tend to keep separated and the number of converged groups roughly equals the number of investors $n$, whereas when $n$ is very large there are more "middle men" to connect the investors and this leads to a smaller number of converged groups. ∎

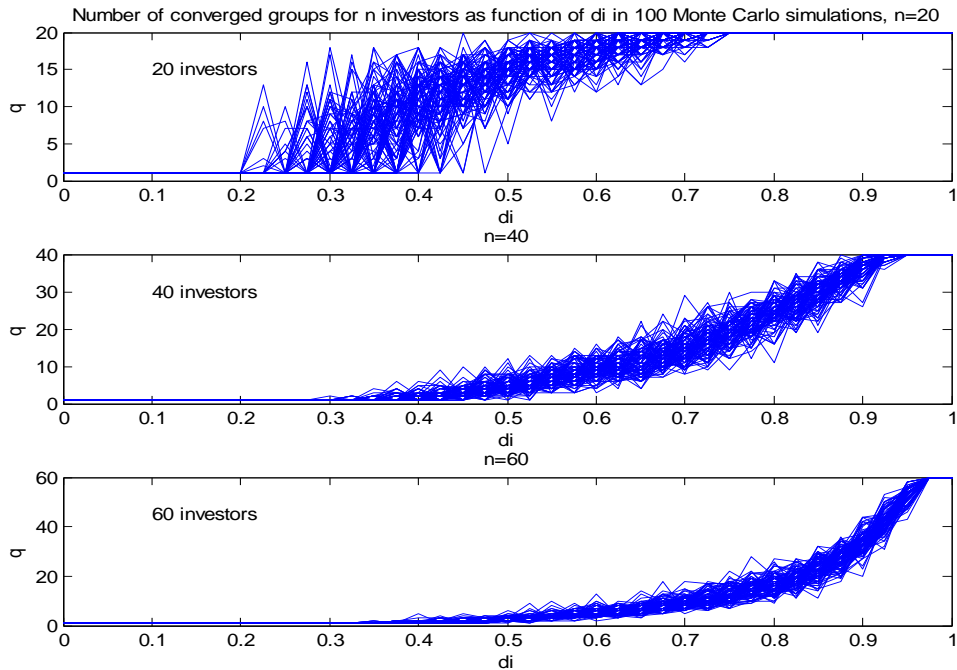

Fig. 4: The number of converged groups of the expected prices $\bar{p}_{i,\infty}$ as function of $d_i$ for $n = 20$ (top), 40 (middle) and 60 (bottom) investors in 100 Monte Carlo simulations with Local Reference.

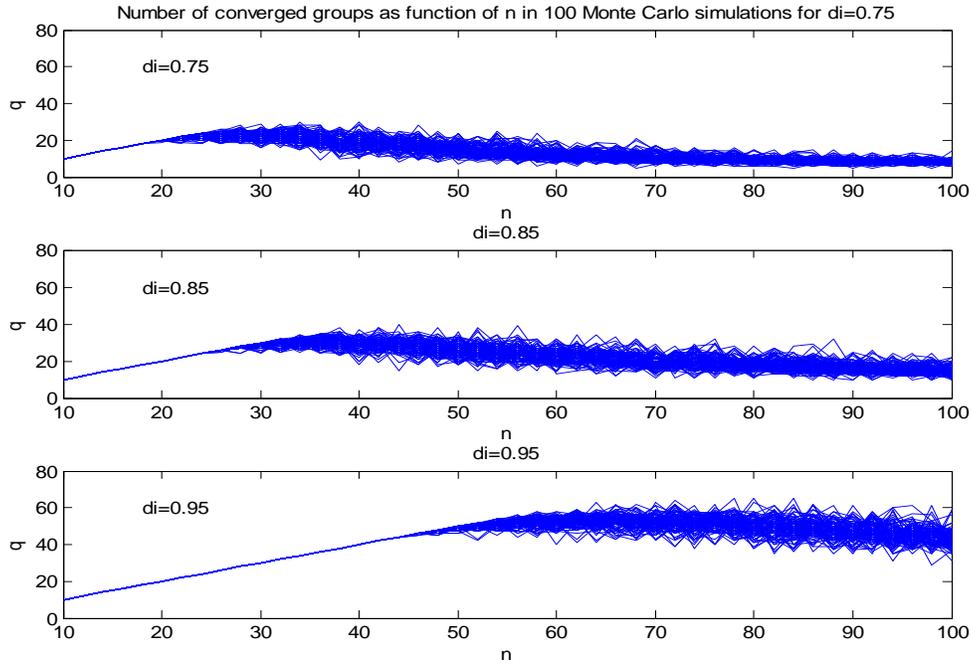

Fig. 5: The number of converged groups of the expected prices $\bar{p}_{i,\infty}$ as function of $n$ for $d_i = 0.75$ (top), 0.85 (middle) and 0.95 (bottom) in 100 Monte Carlo simulations with Local Reference.

**Table 1: Average number of converged groups $\pm$ standard deviation for different values of $d_i$ and $n$ with $b = 1$ and Local Reference.**

| di \ n | 20 | 40 | 60 | 80 | 100 |
|---|---|---|---|---|---|
| 0.2 | 1.08 $\pm$ 0.79 | 1.00 $\pm$ 0.00 | 1.00 $\pm$ 0.00 | 1.00 $\pm$ 0.00 | 1.00 $\pm$ 0.00 |
| 0.3 | 3.86 $\pm$ 3.87 | 1.00 $\pm$ 0.00 | 1.00 $\pm$ 0.00 | 1.00 $\pm$ 0.00 | 1.00 $\pm$ 0.00 |
| 0.4 | 10.04 $\pm$ 4.31 | 1.93 $\pm$ 1.33 | 1.34 $\pm$ 0.49 | 1.20 $\pm$ 0.42 | 1.12 $\pm$ 0.32 |
| 0.5 | 14.09 $\pm$ 2.75 | 5.33 $\pm$ 2.64 | 2.65 $\pm$ 0.90 | 2.54 $\pm$ 0.71 | 2.27 $\pm$ 0.79 |
| 0.6 | 17.22 $\pm$ 1.87 | 9.31 $\pm$ 3.11 | 5.17 $\pm$ 1.41 | 4.34 $\pm$ 1.07 | 3.98 $\pm$ 0.88 |
| 0.7 | 19.40 $\pm$ 0.74 | 15.15 $\pm$ 3.92 | 9.02 $\pm$ 2.31 | 6.84 $\pm$ 1.62 | 6.18 $\pm$ 1.11 |
| 0.8 | 20.00 $\pm$ 0.00 | 23.97 $\pm$ 4.00 | 17.01 $\pm$ 3.75 | 12.53 $\pm$ 2.56 | 10.39 $\pm$ 1.71 |
| 0.9 | 20.00 $\pm$ 0.00 | 36.61 $\pm$ 2.02 | 33.19 $\pm$ 4.25 | 27.77 $\pm$ 3.95 | 23.42 $\pm$ 3.30 |
| 1 | 20.00 $\pm$ 0.00 | 40.00 $\pm$ 0.00 | 60.00 $\pm$ 0.00 | 80.00 $\pm$ 0.00 | 100.00 $\pm$ 0.00 |

**Example 2** (Global Reference): Consider the Stock Price Dynamic Model (39)-(42) with Global Reference (44) and $n = 60$ investors. With parameters $a_i = 0.005$, $\sigma_\varepsilon = 0.02$, $d_i = 0.95$, $b = 0.1$ and initial $p_0 = 10$, $\bar{p}_{i,0}$ ($i = 1, ..., n$) uniformly distributed over the interval [5,25] ($\bar{p}_{i,0} = 5 + 20(\frac{i-1}{n-1}), i = 1, ..., n$) and their uncertainties $\sigma_{i,0}$ ($i = 1, ..., n$) drawn from a random uniform distribution over [0,5], Fig. 6 shows a simulation run of the dynamic model, where the top sub-figure plots the price $p_t$ (heavy line) and the pure random walk price (light line) for comparison, and the middle and bottom sub-figures plot the expected prices $\bar{p}_{i,t}$ ($i = 1, ..., n$) of the $n = 60$ investors and their uncertainties $\sigma_{i,t}$ ($i = 1, ..., n$), respectively. The converged mean price for the simulation run of Fig. 6, as computed from (47) of Theorem 5 (b), is $\bar{p}_\infty = 14.9$. We see from the top sub-figure of Fig. 6, again (similar to the Local Reference case), that the price $p_t$ is moving in a clear up-trend from the initial $p_0 = 10$ towards the final mean price around 14.9 when the investors exchange opinions and reach a consensus as shown in the middle and bottom sub-figures of Fig. 6, while the pure random walk price is just wandering around the initial price $p_0 = 10$ without any trend. Also, we see from Fig. 6 that all investors eventually reach a consensus, i.e., they converge to the same expected price and the same uncertainty, and the convergences are achieved in finite steps, confirming the theoretical results of Theorem 3.

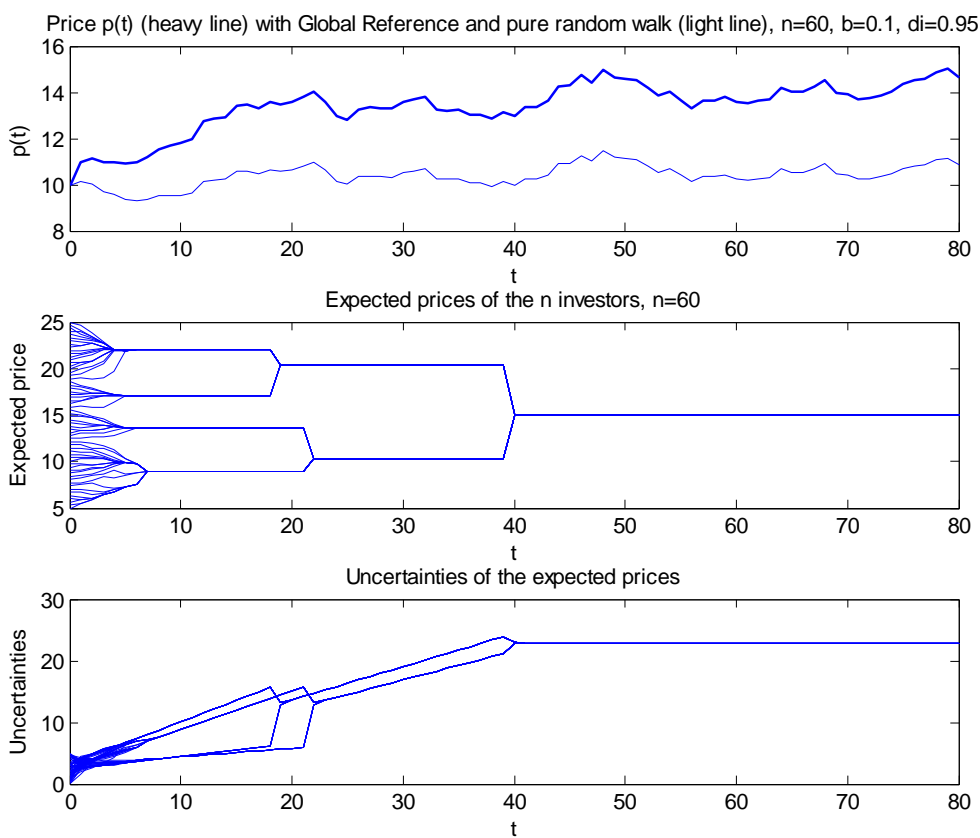

Fig. 6: A simulation run of the Stock Price Dynamic Model (39)-(42) with Global Reference (44) for $n = 60$ investors, where the top sub-figure plots the $p_t$ (heavy line) and pure random walk (light line), and the middle and bottom sub-figures plot the expected prices $\bar{p}_{i,t}$ and their uncertainties $\sigma_{i,t}$ ($i = 1, ..., n$), respectively.

To see the influence of the parameters $d_i$, $n$ and $b$ on the number of steps to reach the consensus, we perform 100 Monte Carlo simulations and the results are summarized as follows: Fig. 7 shows the number of steps to reach consensus as function of $d_i$ for $n = 20, 40$ and $60$ investors (top, middle and bottom sub-figures of Fig. 7, respectively) in the 100 Monte Carlo simulation runs, Fig. 8 shows the number of steps to reach consensus as function of $b$ for $d_i = 0.9, 0.95$ and $0.99$ (top, middle and bottom sub-figures of Fig. 8, respectively) in the 100 Monte Carlo simulations, and Table 2 gives the average number of steps to reach consensus and their standard deviations computed from 100 Monte Carlo simulations for different values of $d_i$ and $b$. We see from Figs. 7 and 8 that the number of steps to reach consensus does not change much for different values of $n$, but is increasing as $d_i$ increases and is decreasing as $b$ increases. The reasons for these properties are that more investors provide more "middle men" to speed up the connection of investors, but more investors need more time to reach consensus and the result is that $n$ does not have much influence on the number of steps to reach consensus, while larger $d_i$ makes it more difficult to find neighbors so that more time is needed to reach consensus because eventually all investors must reach a consensus in this Global Reference case, and larger $b$ speeds up the increase of the uncertainties $\sigma_{i,t}$, leading to a faster rate to form groups and a faster convergence to the consensus. ∎

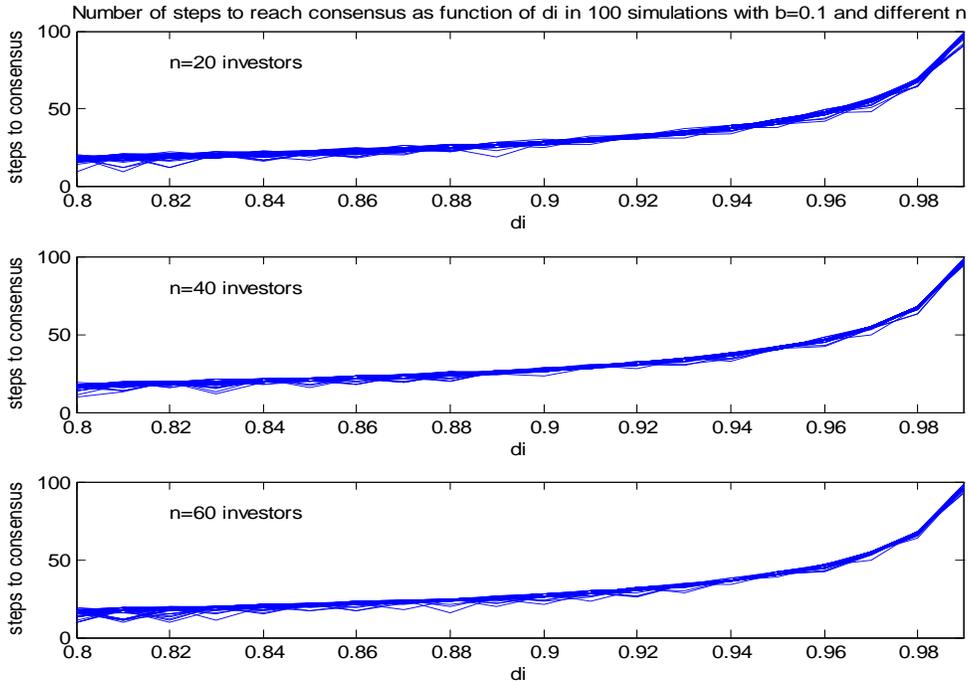

Fig. 7: The number of steps to reach consensus as function of $d_i$ for $n = 20$ (top), $n = 40$ (middle) and $n = 60$ (bottom) investors in 100 Monte Carlo simulations with Global Reference.

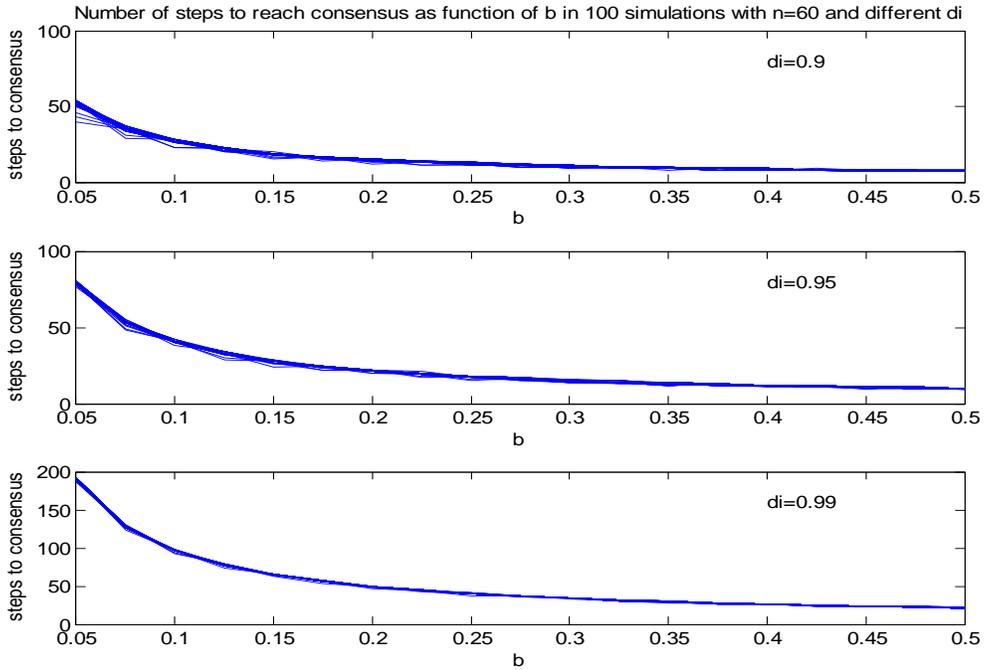

Fig. 8: The number of steps to reach consensus as function of $b$ with $n = 60$ investors for $d_i = 0.9$ (top), $0.95$ (middle) and $0.99$ (bottom) in 100 Monte Carlo simulations with Global Reference.

**Table 2: Average number of steps to reach consensus ± standard deviation for different values of $d_i$ and $b$ with $n = 60$ and Global Reference.**

| b \ di | 0.85 | 0.9 | 0.95 | 0.99 |
|---|---|---|---|---|
| 0.05 | 39.61 ± 1.23 | 51.97 ± 1.53 | 79.16 ± 1.17 | 191.23 ± 1.05 |
| 0.1 | 20.83 ± 0.86 | 27.17 ± 0.73 | 40.76 ± 0.91 | 96.62 ± 1.48 |
| 0.15 | 14.54 ± 0.89 | 18.78 ± 0.57 | 27.99 ± 0.36 | 65.18 ± 0.98 |
| 0.2 | 11.36 ± 0.80 | 14.80 ± 0.46 | 21.78 ± 0.43 | 49.79 ± 0.40 |
| 0.25 | 9.53 ± 0.60 | 12.16 ± 0.46 | 17.88 ± 0.35 | 40.03 ± 0.22 |
| 0.3 | 8.29 ± 0.45 | 10.68 ± 0.52 | 15.00 ± 0.24 | 34.00 ± 0.11 |
| 0.35 | 7.61 ± 0.50 | 9.31 ± 0.50 | 13.25 ± 0.45 | 29.44 ± 0.51 |
| 0.4 | 6.97 ± 0.17 | 8.49 ± 0.49 | 11.94 ± 0.27 | 26.00 ± 0.20 |
| 0.45 | 6.46 ± 0.49 | 7.98 ± 0.14 | 11.00 ± 0.20 | 23.43 ± 0.49 |

**Example 3** (Real Price Reference): Consider the stock price dynamic model (39)-(42) with Real Price Reference (45) and $n = 60$ investors. With parameters $a_i = 0.005$, $\sigma_\varepsilon = 0.02$, $d_i = 0.95$, $b = 0.1$ and initial $p_0 = 10$, $\bar{p}_{i,0}$ ($i = 1, ..., n$) uniformly distributed over the interval [5,25] and their uncertainties $\sigma_{i,0}$ drawn from a random uniform distribution over [0,5], Fig. 9 shows a simulation run of the dynamic model, where the top sub-figure plots the price $p_t$ (heavy line) and the pure random walk price (light line), and the middle and bottom sub-figures plot the expected prices $\bar{p}_{i,t}$ ($i = 1, ..., n$) of the $n = 60$ investors and their uncertainties $\sigma_{i,t}$, respectively. The converged mean price for the simulation run in the top sub-figure of Fig. 9 is about 13 which is much less than the converged expected price $\bar{p}_\infty = 15$ shown in the middle sub-figure of Fig. 9, confirming the theoretical result of Theorem 5 (c). Also, the middle and bottom sub-figures of Fig. 9 confirm the theoretical results of Theorem 4 that the expected prices converge to a constant consensus $\bar{p}_\infty$ and the uncertainties reach the same $\sigma_t$ in finite time, but this $\sigma_t$ keeps increasing and has no sign to converge to a constant.

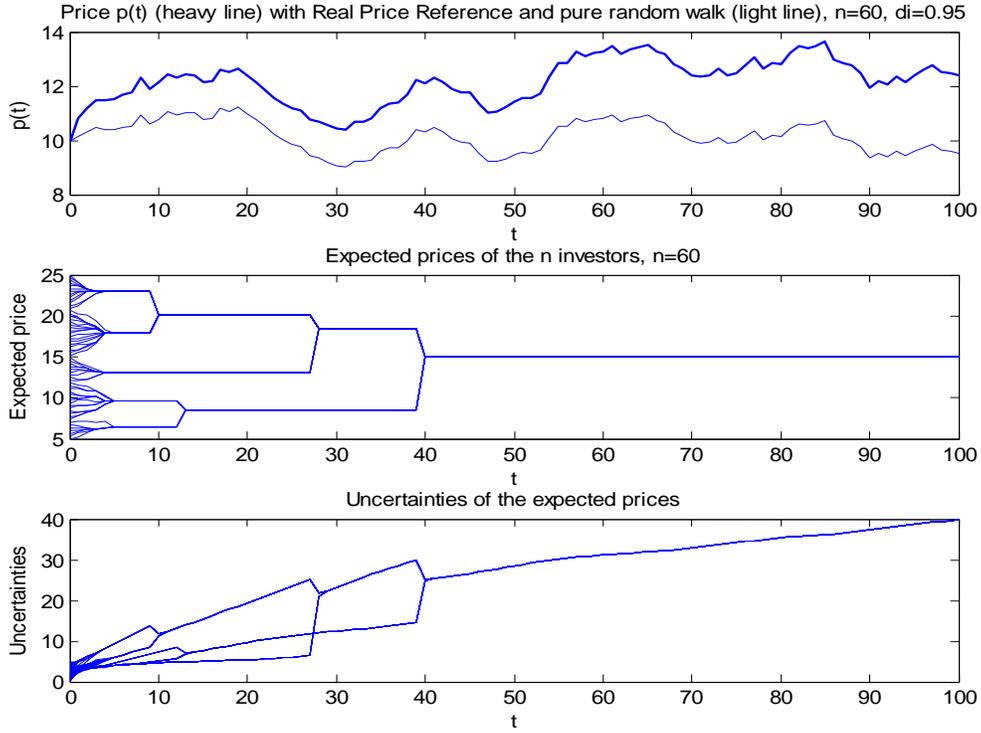

Fig. 9: A simulation run of the Stock Price Dynamic Model (39)-(42) with Real Price Reference (45) for $n = 60$ investors, where the top sub-figure plots the $p_t$ (heavy line) and pure random walk (light line), and the middle and bottom sub-figures plot the expected prices $\bar{p}_{i,t}$ and their uncertainties $\sigma_{i,t}$ ($i = 1, ..., n$), respectively.

To see the influence of the parameters $d_i$, $n$ and $b$ on the number of steps for the expected prices $\bar{p}_{i,t}$ to reach the consensus $\bar{p}_\infty$, we perform 100 Monte Carlo simulations and the results are as follows: Fig. 10 shows the number of steps to reach consensus as function of $d_i$ for $n = 20, 40$ and $60$ investors (top, middle and bottom sub-figures of Fig. 10, respectively) in the 100 Monte Carlo simulation runs, Fig. 11 shows the number of steps to reach consensus as function of $b$ for $d_i = 0.9, 0.95$ and $0.99$ (top, middle and bottom sub-figures of Fig. 11, respectively) in the 100 Monte Carlo simulations, and Table 3 gives the average number of steps to reach consensus and their standard deviations computed from 100 Monte Carlo simulations for different values of $d_i$ and $b$. We see from Figs. 10 and 11 that, similar to the Global Reference case, the number of steps to reach consensus does not change much for different values of $n$, but is increasing as $d_i$ increases and is decreasing as $b$ increases. The reasons for such behaviors are also the same as those for the Global Reference case. The main difference between the Real

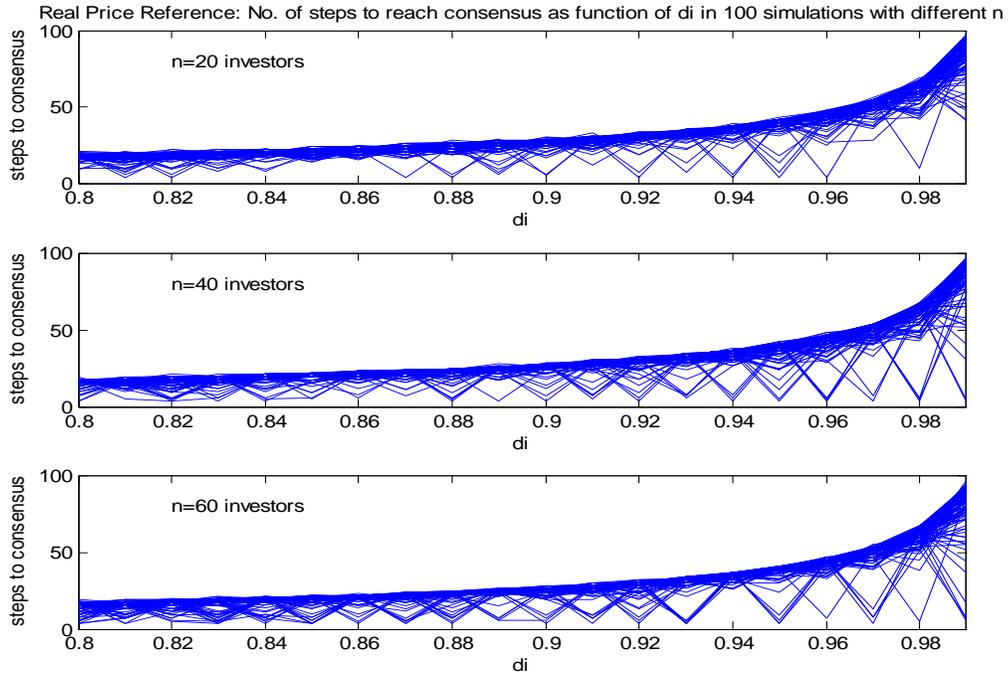

Fig. 10: The number of steps to reach consensus as function of $d_i$ for $n = 20$ (top), 40 (middle) and 60 (bottom) investors in 100 Monte Carlo simulations with Real Price Reference.

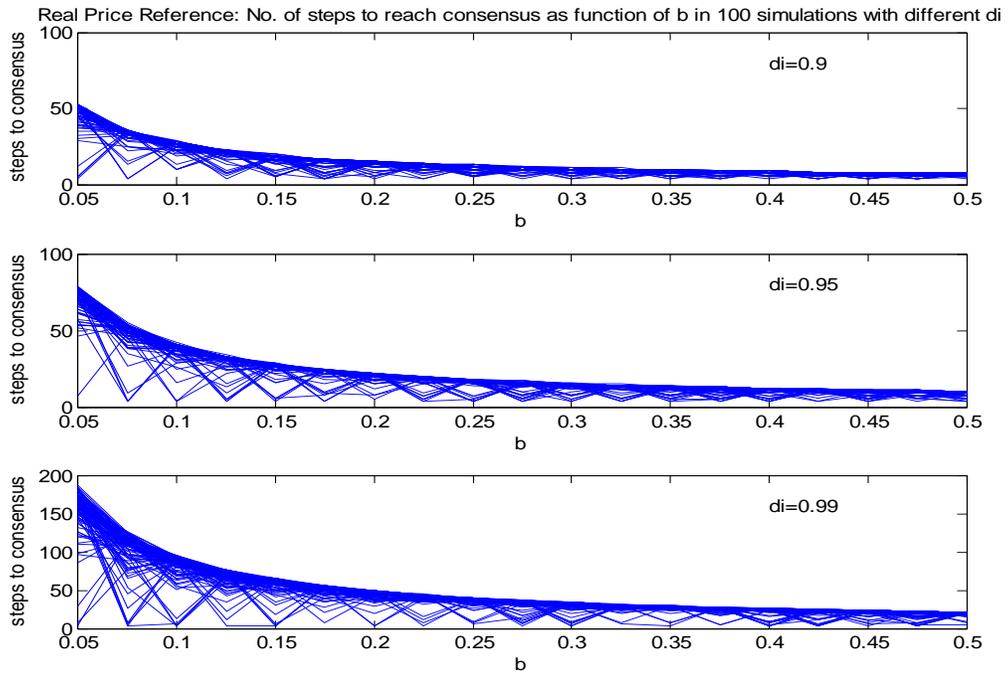

Fig. 11: The number of steps to reach consensus as function of $b$ with $n = 60$ investors for $d_i = 0.9$ (top), 0.95 (middle) and 0.99 (bottom) in 100 Monte Carlo simulations with Real Price Reference.

**Table 3: Average number of steps to reach consensus ± standard deviation for different values of $d_i$ and $b$ with $n = 60$ and Real Price Reference.**

| b \ di | 0.85 | 0.9 | 0.95 | 0.99 |
|---|---|---|---|---|
| 0.05 | 35.99 ± 5.80 | 47.25 ± 8.71 | 71.28 ± 11.74 | 153.24 ± 36.05 |
| 0.1 | 18.76 ± 3.75 | 23.44 ± 6.53 | 37.26 ± 5.27 | 80.82 ± 20.24 |
| 0.15 | 12.56 ± 2.74 | 16.65 ± 3.92 | 25.18 ± 5.35 | 55.49 ± 12.44 |
| 0.2 | 10.32 ± 1.66 | 13.66 ± 2.35 | 19.95 ± 3.51 | 44.42 ± 8.48 |
| 0.25 | 8.28 ± 1.57 | 10.82 ± 2.17 | 16.51 ± 2.72 | 34.76 ± 8.19 |
| 0.3 | 7.53 ± 0.95 | 9.25 ± 1.93 | 13.61 ± 2.90 | 30.37 ± 6.31 |
| 0.35 | 6.93 ± 0.86 | 8.55 ± 1.13 | 12.50 ± 1.57 | 26.35 ± 4.91 |
| 0.4 | 6.41 ± 0.70 | 7.80 ± 0.84 | 11.14 ± 1.63 | 24.07 ± 4.15 |
| 0.45 | 6.04 ± 0.58 | 7.27 ± 0.96 | 10.10 ± 1.64 | 21.48 ± 4.13 |

Price Reference and the Global Reference results is, by comparing Figs. 7 and 8 with Figs. 10 and 11 and Table 2 with Table 3, that the variance of the number of steps to reach consensus in the Real Price Reference case is much larger than that in the Global Reference case, and the reason for this difference is that more randomness is introduced through the random prices, leading to more variance for the convergence steps. ∎

**Example 4** (Price dynamics with Followers): Figs. 12, 13 and 14 illustrate simulation runs of the Stock Price Dynamic Model with Followers in Section IV for the Local, Global and Real Price Reference indicators (50), (51) and (52), respectively, where the top sub-figures plot the prices $p_t$ (heavy lines) and pure random walk prices (light lines), and the middle and bottom sub-figures plot the expected prices $\bar{p}_{i,t}$ ($i = 1, ..., n$) and their uncertainties $\sigma_{i,t}$ ($i = 1, ..., n$), respectively. The parameter setting is: $n = 60$, $a_i = 0.002$ and $\sigma_\varepsilon = 0.02$ for all Figs. 12-14, $c_i = 0.0001$, $d_i = 0.6$ and $b = 1$ for Fig. 12, $c_i = 0.1$, $d_i = 0.95$ and $b = 0.1$ for Figs. 13 and 14. The initial conditions are: $p_0 = 10$ and $\bar{p}_{i,0}$ ($i = 1, ..., n$) are uniformly distributed over the interval [5,25] for all Figs. 12-14, and the uncertainties $\sigma_{i,0}$ are drawn from a random uniform distribution over (0,1) for Fig. 12 and (0,5) for Figs. 13 and 14.

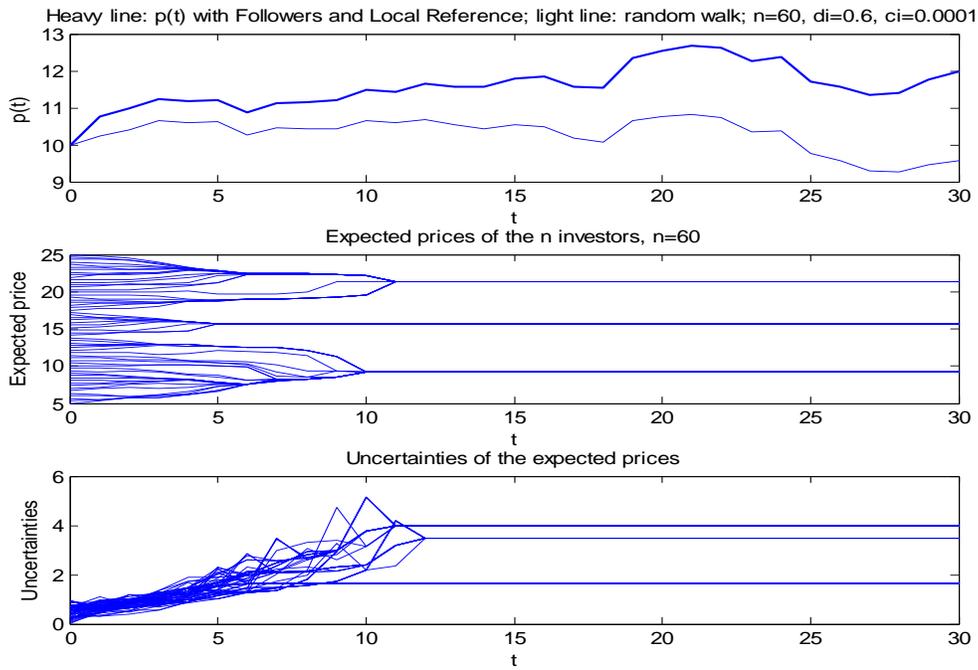

Fig. 12: A simulation run of the Stock Price Dynamic Model with Followers and Local Reference indicator (50) for $n = 60$ investors, where the top sub-figure plots the $p_t$ (heavy line) and pure random walk (light line), and the middle and bottom sub-figures plot the expected prices $\bar{p}_{i,t}$ and their uncertainties $\sigma_{i,t}$ $(i = 1, \dots, n)$, respectively.

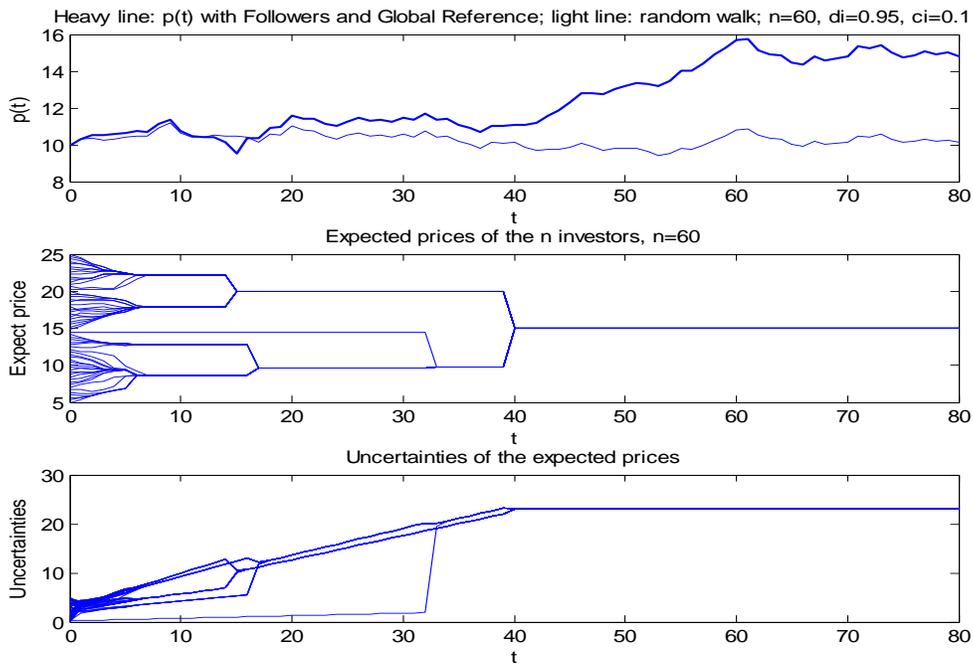

Fig. 13: A simulation run of the Stock Price Dynamic Model with Followers and Global Reference indicator (51).

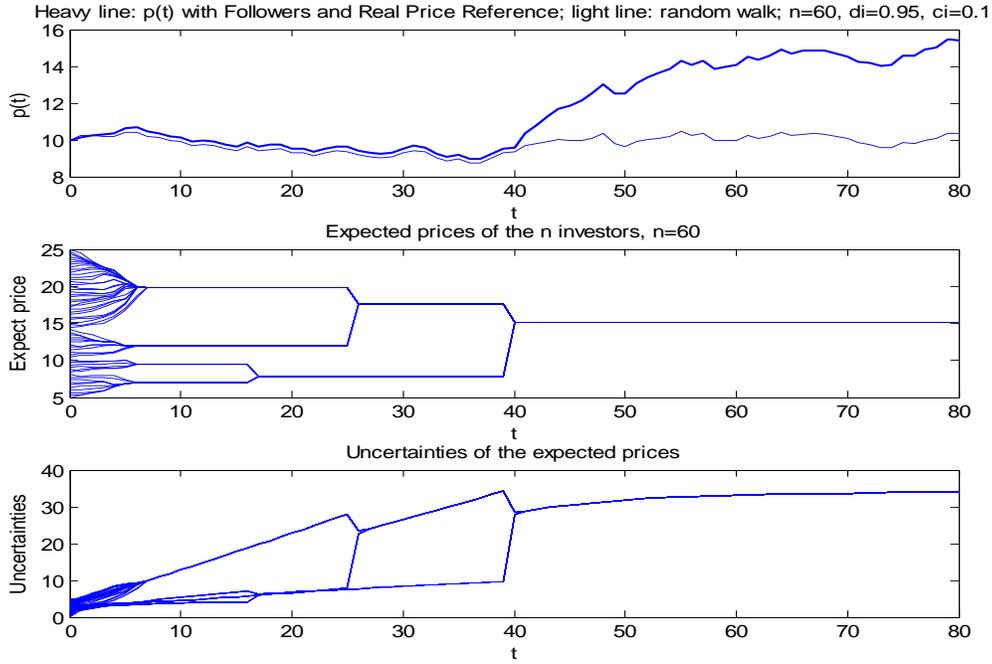

Fig. 14: A simulation run of the Stock Price Dynamic Model with Followers and Real Price Reference indicator (52).

From Figs. 12-14 we see that for the Local Reference case (Fig. 12) the price $p_t$ moves away from the random walk (i.e. starts trending) in the very early stage, but for the Global and Real Price Reference cases (Figs. 13 and 14) the prices $p_t$ are very close to random walk in the early stages and start trending (move away from random walk) only when all the investors reach a consensus. The reason for such behavior is that in the Local Reference case it is easier for an investor to become a follower because the investor only needs to follow the average of his neighbors, whereas in the Global and Real Price Reference cases an investor must follow the average of all the investors or the same real price to become a follower, which are more difficult conditions to satisfy and the result is that we see much less followers in the Global and Real Price Reference cases to move the price away from random walk before all the investors reach the consensus, after which all investors become followers and the price starts moving away from random walk very quickly. ∎

**Example 5** (Price dynamics with Contrarians): Figs. 15, 16 and 17 illustrate simulation runs of the Stock Price Dynamic Model with Contrarians in Section IV for the Local, Global and Real Price Reference indicators (53), (54) and (55), respectively, where the top sub-figures plot the prices $p_t$ (heavy lines) and pure random walk prices (light lines), and the middle and bottom sub-figures plot the expected prices $\bar{p}_{i,t}$ ($i = 1, ..., n$) and their uncertainties $\sigma_{i,t}$, respectively. The parameter setting is: $n = 60$, $a_i = 0.002$ and $\sigma_\varepsilon = 0.02$ for all Figs. 15-17, $c_i = 0.01$, $d_i = 0.6$ and $b = 1$ for Fig. 15, $c_i = 0.2$, $d_i = 0.95$ and $b = 0.1$ for Figs. 16 and 17. The initial conditions are: $p_0 = 10$ and $\bar{p}_{i,0}$ ($i = 1, ..., n$) are uniformly distributed over the interval [5,25] for all Figs. 15-17, and the uncertainties $\sigma_{i,0}$ are drawn from a random uniform distribution over (0,1) for Fig. 15 and (0,5) for Figs. 16 and 17. From Figs. 15-17 we see that the prices $p_t$ are quite different from random walk during the early stage before the investors reach consensus and follow the random walks closely after all investors converge to their final opinions. The reason for such behavior is that contrarians are easier to appear when investors' opinions are quite different from each other during the early stage, while as investors exchange opinions and approach consensus locally or globally there is less chance for contrarians to appear that leads to $I_{t,i} = 0$ and the price equation (49) is reduced to random walk $\ln(p_{t+1}) = \ln(p_t) + \varepsilon_t$. ∎

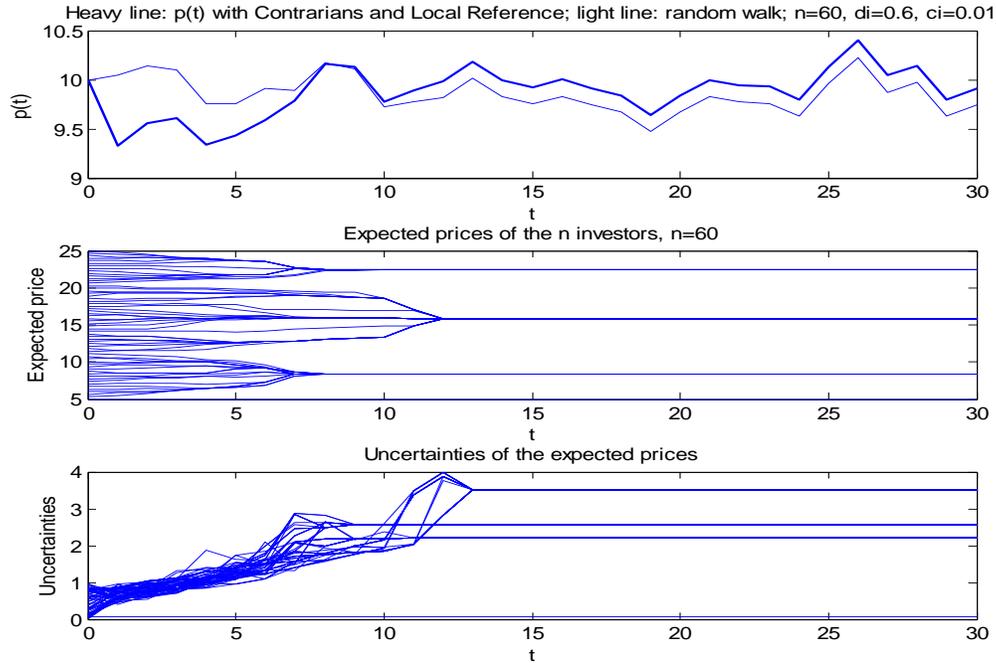

Fig. 15: A simulation run of the Stock Price Dynamic Model with Contrarians with Local Reference indicator (53).

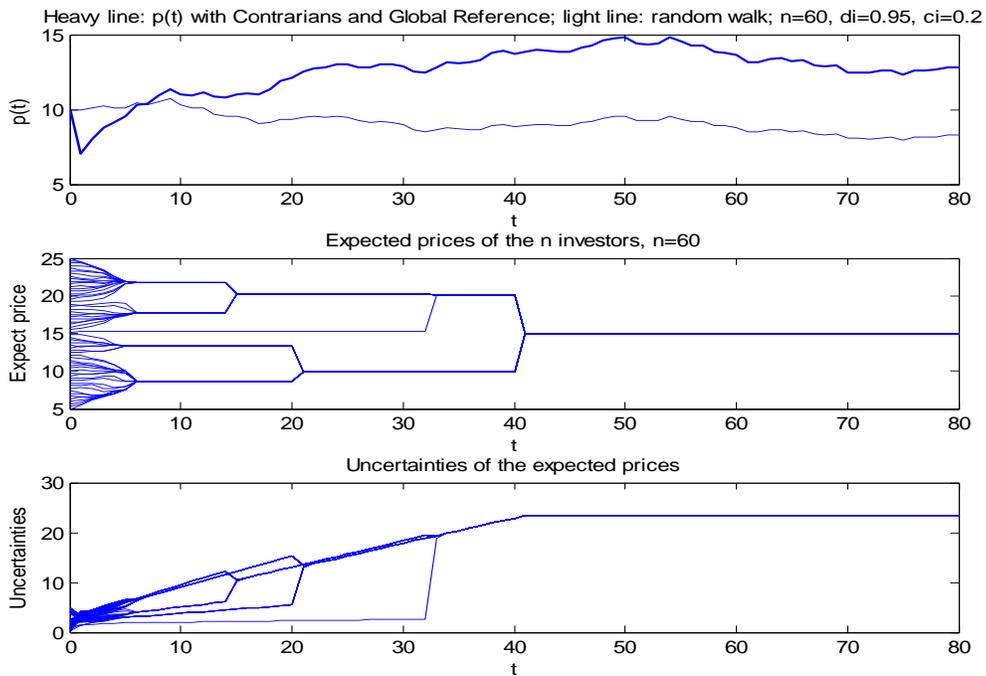

Fig. 16: A simulation run of the Stock Price Dynamic Model with Contrarians and Global Reference indicator (54).

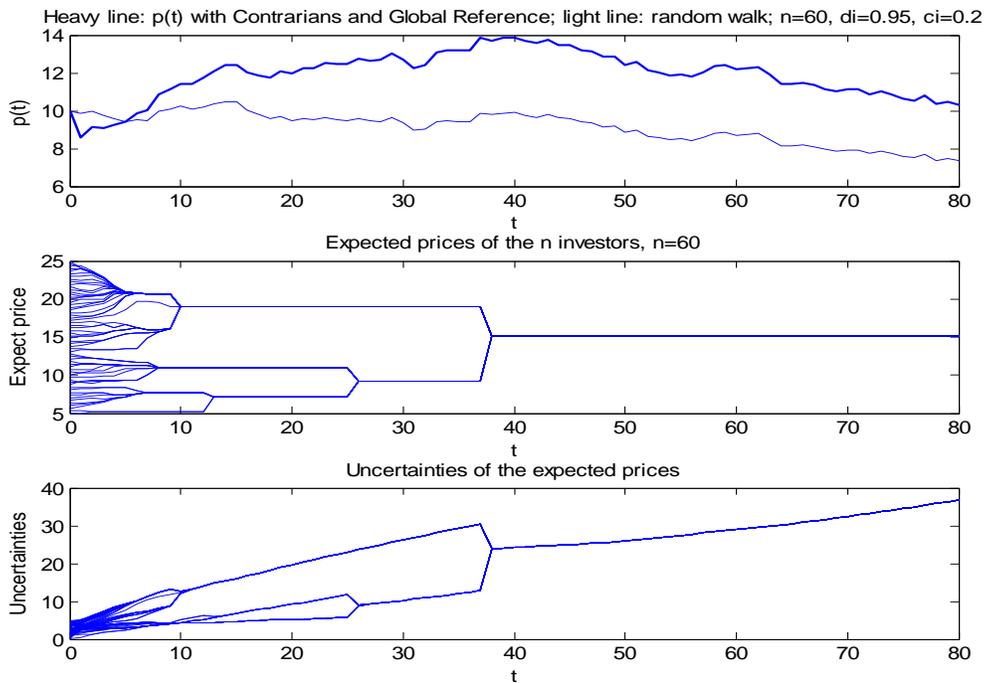

Fig. 17: A simulation run of the Stock Price Dynamic Model with Contrarians and Real Price Reference indicator (55).

**Example 6** (Price dynamics with Manipulators): Simulation results of the Stock Price Dynamic Model with Manipulators are shown in Fig. 18 with Local Reference (43), in Figs. 19 and 20 with the Global Reference (44), and in Figs. 21 and 22 with Real Price Reference (45), where the top sub-figures plot the prices $p_t$ (heavy lines) and pure random walk prices (light lines), and the middle and bottom sub-figures plot the expected prices $\bar{p}_{i,t}$ ($i = 1, \ldots, n$) and their uncertainties $\sigma_{i,t}$, respectively. The parameter setting is: $n = 60$, $a_i = 0.002$ and $\sigma_\varepsilon = 0.02$ for all Figs. 18-22, $b = 1$ and $d_i = 0.6$ for all $i = 1, \ldots, n$ except $d_{50} = 1$ for Fig. 18, $b = 0.1$ and $d_i = 0.95$ for all $i = 1, \ldots, n$ except $d_{50} = 1$ for Figs. 19 and 21, and $b = 0.1$, $d_i = 0.95$ for all $i = 1, \ldots, n$ except $d_{40} = d_{50} = 1$ for Figs. 20 and 22. The initial conditions are: $p_0 = 10$ and $\bar{p}_{i,0}$ ($i = 1, \ldots, n$) are uniformly distributed over the interval [5,25] for all Figs. 18-22, and the uncertainties $\sigma_{i,0}$ are drawn from a random uniform distribution over (0,1) for Fig. 18 and (0,5) for Figs. 19-22.

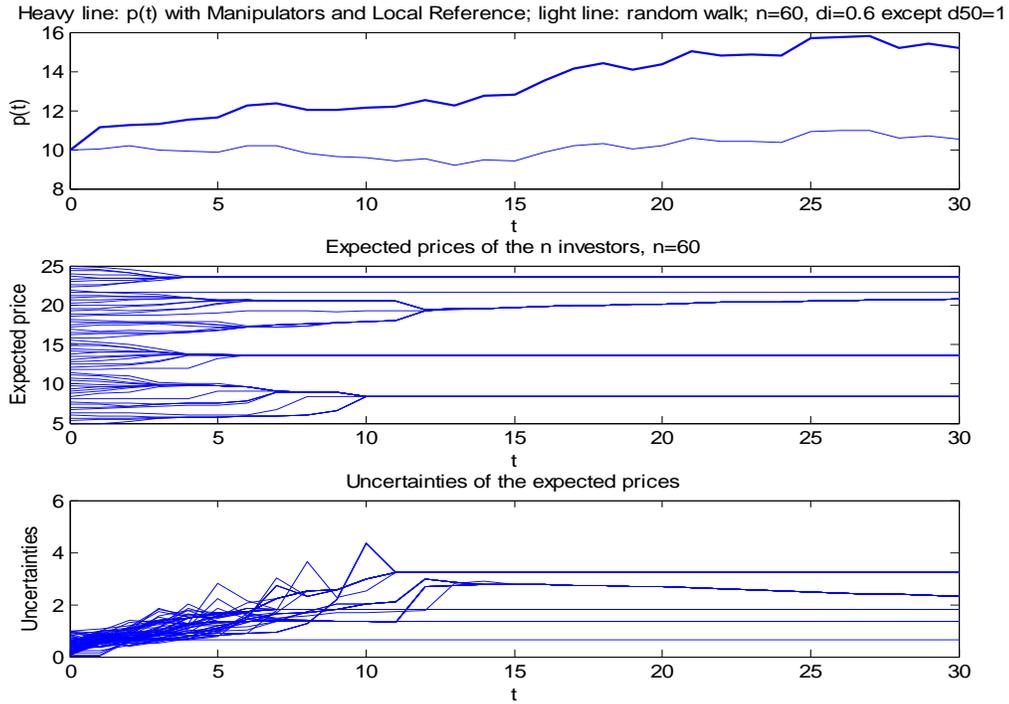

Fig. 18: A simulation run of the Stock Price Dynamic Model with Manipulators and Local Reference (43) for $n = 60$ investors with one manipulator $d_{50} = 1$.

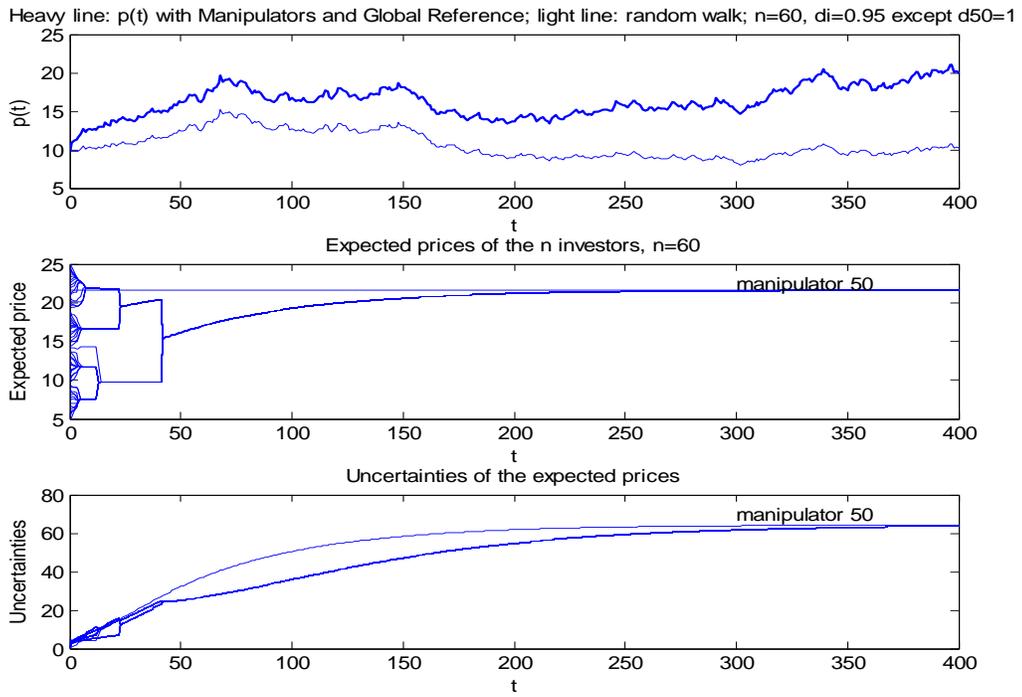

Fig. 19: A simulation run of the Stock Price Dynamic Model with Manipulators and Global Reference (44) for $n = 60$ investors with one manipulator $d_{50} = 1$.

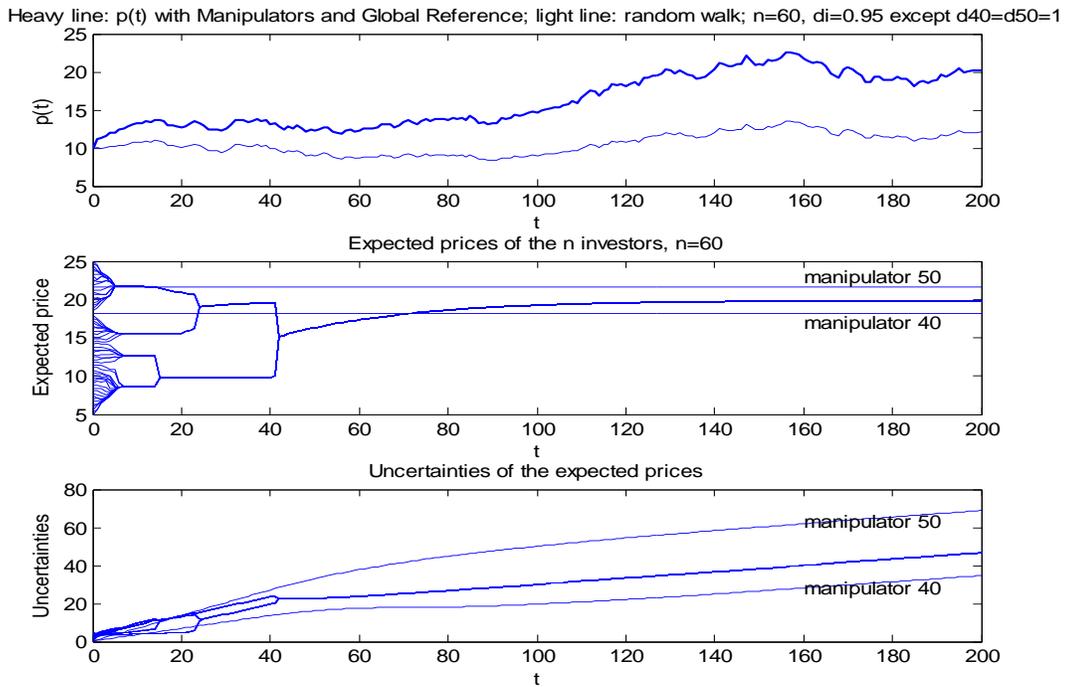

Fig. 20: A simulation run of the Stock Price Dynamic Model with Manipulators and Global Reference (44) for $n = 60$ investors with two manipulator $d_{40} = d_{50} = 1$.

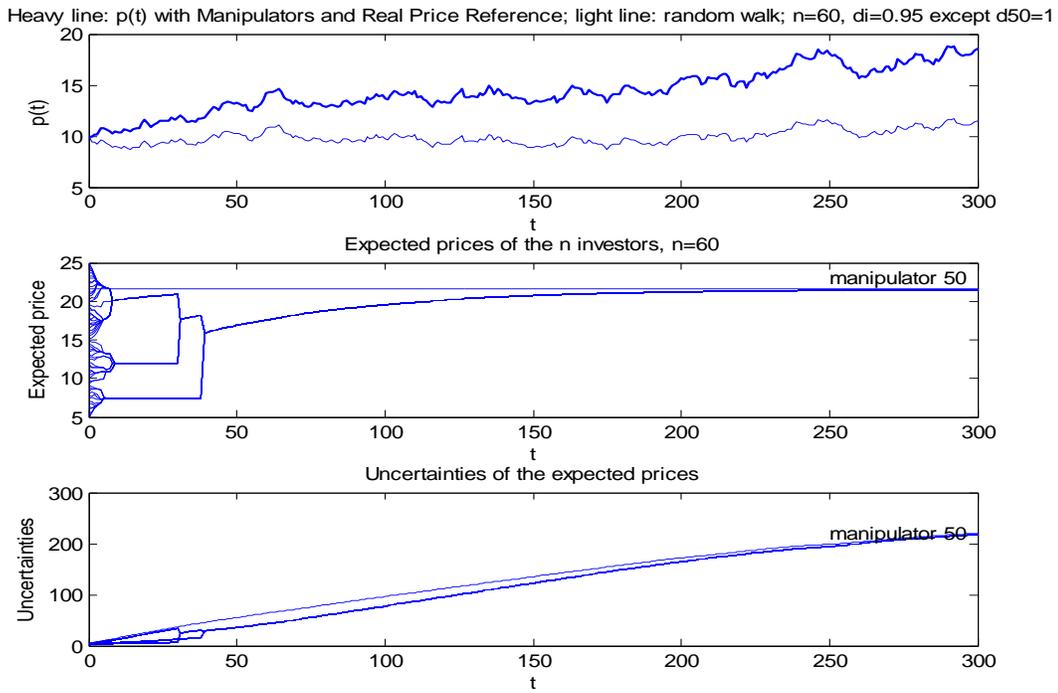

Fig. 21: A simulation run of the Stock Price Dynamic Model with Manipulators and Real Price Reference (45) for $n = 60$ investors with one manipulator $d_{50} = 1$.

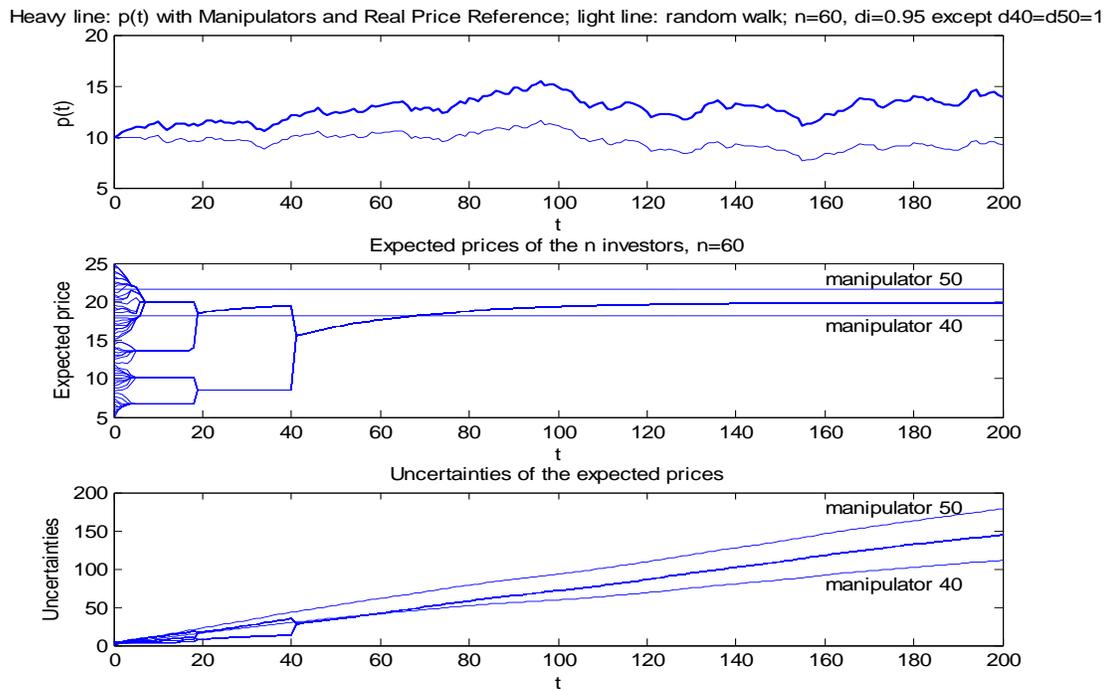

Fig. 22: A simulation run of the Stock Price Dynamic Model with Manipulators and Real Price Reference (45) for $n = 60$ investors with two manipulator $d_{40} = d_{50} = 1$.

We see from Fig. 18 that the manipulator $d_{50} = 1$ does not have much influence in the Local Reference case because investors in this case view the average of their neighbors' opinions as the reference signal so that the manipulator can only influence a very limited number of investors. For the Global Reference cases in Figs. 19 and 20 we see that when there is only one manipulator $d_{50} = 1$ (Fig. 19), all investors' expected prices converge to the manipulated value and their uncertainties converge to a constant, but when there are two manipulators $d_{40} = d_{50} = 1$ (Fig. 20), the ordinary investors' expected prices converge to the average of the two manipulators' expected prices and the uncertainties of all investors, including the two manipulators, go to infinity. The reason for the divergence of the uncertainties is that when there are more than one manipulators, each manipulator cannot completely control the situation so that their uncertainty inputs $u_i(t+1) = b \left| \bar{p}_{i,t} - \frac{1}{n} \sum_{j=1}^{n} \bar{p}_{j,t} \right|$ do not converge to zero that push their uncertainties to infinity. The Real Price Reference results Figs. 21 and 22 are similar to the Global Reference results in Figs. 19 and 20, except that even when there is only one manipulator (Fig. 21), the uncertainties of all investors, including the manipulator, go to infinity because even the manipulator does not know the real prices so that the manipulator's uncertainty also goes to infinity along with the uncertainties of the ordinary investors. ∎

## VI. Application to Hong Kong Stocks

In the real stock markets, the only information we have is the price data $p_t, t = 0,1,2,\dots$ Suppose we use the Stock Price Dynamic Model (39)-(45) to model a real stock price series $p_t$ ($t = 0,1,2,\dots$), then building the model (39)-(45) boils down to identifying the model parameters: $n$ (the number of investors), $a_i$ (the relative strength of investor $i$), $d_i$ (the confidence bound of investor $i$), $b$ (the scaling parameter for uncertainty inputs) and $\sigma_\varepsilon$ (the standard deviation of the noise term), based on the information set $\{p_t: t = 0,1,2,\dots\}$. This is a very hard problem because these parameters influence the price $p_t$ through the very nonlinear channels in (39)-(45). Although some general optimization methods such as the genetic algorithms may be used to estimate these parameters based on the price data, we leave it to future research. Here in this paper we propose some aggregated variables that combine the expected prices and their uncertainties of all the investors to simplify the model identification problem. Specifically, we define the combined expected price and the combined uncertainty as follows:

**Definition 5:** Consider the Stock Price Dynamic Model (39)-(45). The *combined uncertainty* $\sigma_t$ is defined as:

$$\sigma_t = \frac{1}{\sum_{i=1}^{n}\left(\frac{a_i}{\sigma_{i,t}}\right)} \tag{60}$$

and the *combined expected price* $\bar{p}_t$ is defined as:

$$\bar{p}_t = e^{\sigma_t \sum_{i=1}^{n}\left(\frac{a_i \ln(\bar{p}_{i,t})}{\sigma_{i,t}}\right)} \tag{61}$$

∎

With $\sigma_t$ and $\bar{p}_t$ defined in (60) and (61), the price dynamic equation (39) becomes:

$$\ln(p_{t+1}) = \ln(p_t) + \frac{\ln(\bar{p}_t) - \ln(p_t)}{\sigma_t} + \varepsilon_t \tag{62}$$

which can be verified by substituting (60) and (61) into (62) to get back (39). Comparing (62) with (39) we see that there are $n$ individual investors in (39) while there is only one pseudo-investor in (62) whose net effect is equivalent to the combination of the $n$ individual investors in (39). In fact, the combined uncertainty (60) and the combined expected price (61) are chosen in such a way that the sum $\sum_{i=1}^{n} \frac{a_i[\ln(\bar{p}_{i,t}) - \ln(p_t)]}{\sigma_{i,t}}$ in (39) is reduced to the single term $\frac{\ln(\bar{p}_t) - \ln(p_t)}{\sigma_t}$ in (62). From (62) we see that the $\bar{p}_t$ may be viewed as the overall expected price from the $n$ investors with the $\sigma_t$ being the overall uncertainty. Now instead of estimating the parameters $n$, $a_i$, $d_i$, $b$ and $\sigma_\varepsilon$ in the price dynamic model (39)-(45), our task is to estimate the combined expected price $\bar{p}_t$ and the combined uncertainty $\sigma_t$ based on the priced data $\{p_t: t = 0,1,2,...\}$. Let $r_{t+1} = \ln(p_{t+1}/p_t)$, $s_t = (1, -\ln(p_t))^T$ and $v_t = \left(\frac{1}{\sigma_t}\ln(\bar{p}_t), \frac{1}{\sigma_t}\right)^T$, we obtain from (62) that

$$r_{t+1} = s_t^T v_t + \varepsilon_t \tag{63}$$

Comparing with the fast-changing price $p_t$, we can assume that the investors' expected prices $\bar{p}_{i,t}$ and their uncertainties $\sigma_{i,t}$ are slowly time-varying, therefore we view $v_t$ as a slowly time-varying parameter vector. A good method to estimate slowly time-varying parameters is the standard Recursive Lease Squares Algorithm with Exponential Forgetting which minimizes the weighted summation of error:

$$E_{t+1}(v) = \sum_{i=1}^{t} \lambda^{t-i} (r_{i+1} - s_i^T v)^2 \tag{64}$$

to obtain the estimate of $v_t$, denoted as $\hat{v}_t = (\hat{v}_{1t}, \hat{v}_{2t})^T$, through the following recursive computations (see, e.g., page 53 of [2]):

$$\hat{v}_t = \hat{v}_{t-1} + K_t(r_{t+1} - s_t^T \hat{v}_{t-1}) \tag{65}$$

$$K_t = \frac{P_{t-1} s_t}{(s_t^T P_{t-1} s_t + \lambda)} \tag{66}$$

$$P_t = (I - K_t s_t^T) P_{t-1} / \lambda \tag{67}$$

where $\lambda \in (0,1)$ is a forgetting factor to put more weights on recent data. With $\hat{v}_t = (\hat{v}_{1t}, \hat{v}_{2t})^T$ computed from the algorithm (65)-(67), we obtain the estimates of the combined uncertainty $\sigma_t$ and the combined expected price $\bar{p}_t$ as:

$$\hat{\sigma}_t = \frac{1}{\hat{v}_{2t}} \tag{68}$$

$$\widehat{\bar{p}}_t = e^{\hat{v}_{1t}/\hat{v}_{2t}} \tag{69}$$

We now apply the algorithm (65)-(69) to the daily closing prices[5] of fifteen major banking and real estate stocks listed in the Hong Kong Stock Exchange for the recent two-year period from Dec. 5, 2013 to Dec. 4, 2015; Figs. 23-37 show the results with $\lambda = 0.999$ and the initial $\hat{v}_0 = (0.5, 0.1)^T$ and $P_0 = \begin{pmatrix} 10 & 0 \\ 0 & 10 \end{pmatrix}$ for all the stocks, where the top sub-figures of Figs. 23-37 plot the daily closing prices of the fifteen stocks from Dec. 5, 2013 to Dec. 4, 2015 (492 data points) which are used as the $p_t$ in the algorithm (65)-(69), and the middle and bottom sub-figures of Figs. 23-37 plot the estimated combined expected prices $\widehat{\bar{p}}_t$ and the estimated combined uncertainties $\hat{\sigma}_t$, respectively, for the fifteen stocks. From Figs. 23-37 we observe the following:

*Observation 1*: The estimated combined expected prices $\widehat{\bar{p}}_t$ were generally smooth and did not change rapidly during a short period of time.

*Observation 2*: The estimated combined uncertainties $\hat{\sigma}_t$ were mostly smooth but changed rapidly during some periods of time; when the $\hat{\sigma}_t$ increased rapidly, the current trend of the stock price would be reversed in the near future, whereas when the $\hat{\sigma}_t$ decreased rapidly, the current trend of the stock price would continue in the near future.

---

[5] All stock price data used in this paper were downloaded from http://finance.yahoo.com and were adjusted for dividends and splits.

An explanation of Observation 2 is the following: A rapid increase of the combined uncertainty implies that the investors overall do not agree with the current price behavior so that a trend reversal is expected in the near future, whereas a rapid decrease of the combined uncertainty indicates that the investors overall agree with the current price trend so that it will continue in the near future. To see how Observation 2 is obtained, we now analyze some of Figs. 23-37 as follows (more discussions are given in the captions below Figs. 23-37):

Figs. 24 (China Constr. Bank) and 25 (Ind. & Com. Bank of China): A rapid increase of the combined uncertainty in Circle 1 revealed the disagreement of the investors with the new high of the price in Circle 1 so that the price declined sharply after a short period. In Circle 2, the clear decrease of the combined uncertainty indicated that the investors agreed with the sharp decline of price in Circle 2 so that the trend continued in the near future.

Figs. 29 (BOC Hong Kong) and 30 (Hang Seng Bank): A steady big increase of the combined uncertainty in Circle 1 indicated the increasing disagreement of the investors with the big price rise in Circle 1 so that a sharp price decline was expected in the near future which indeed occurred.

Figs. 33 (New World Develop.) and 35 (The Bank of East Asia): The big increases of the combined uncertainty in Circles 1 and 2 indicated the disagreement of the investors with the sharp price rise in Circle 1 and the big price drop in Circle 2 so that the price declined after Circle 1 and bounded back after Circle 2 when the investors took actions to show their disagreement with the price behaviors in Circles 1 and 2.

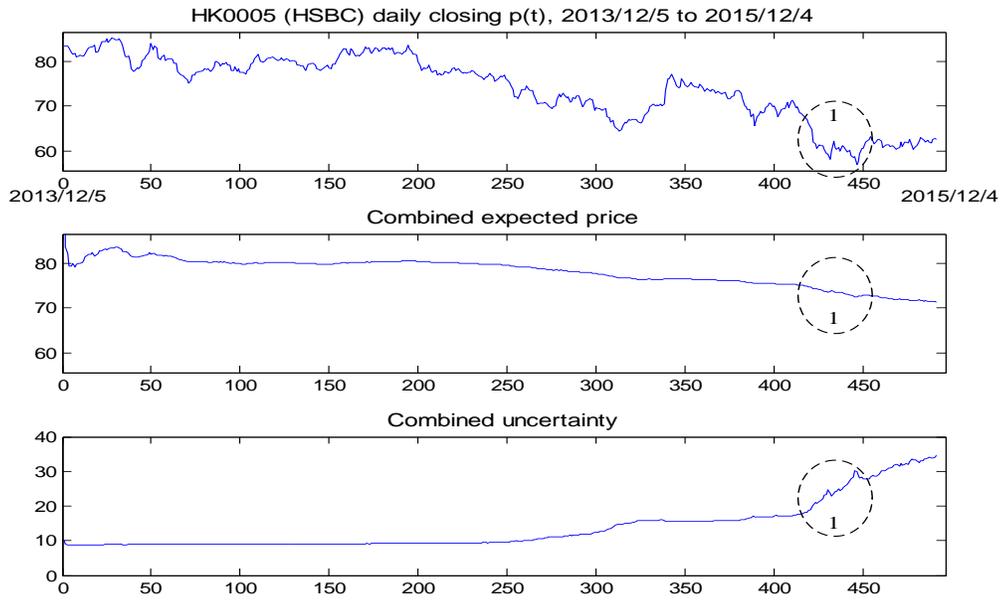

Fig. 23: Top: Daily closing prices $p_t$ of HK0005 (HSBC) from Dec. 5, 2013 to Dec. 4, 2015 (492 data points). Middle: Estimated combined expected prices $\hat{\bar{p}}_t$. Bottom: Estimated combined uncertainties $\hat{\sigma}_t$. The sharp increase of $\hat{\sigma}_t$ in Circle 1 indicates that the investors do not agree with the sharp decline of the price in Circle 1 so that the price down trend is expected to stop.

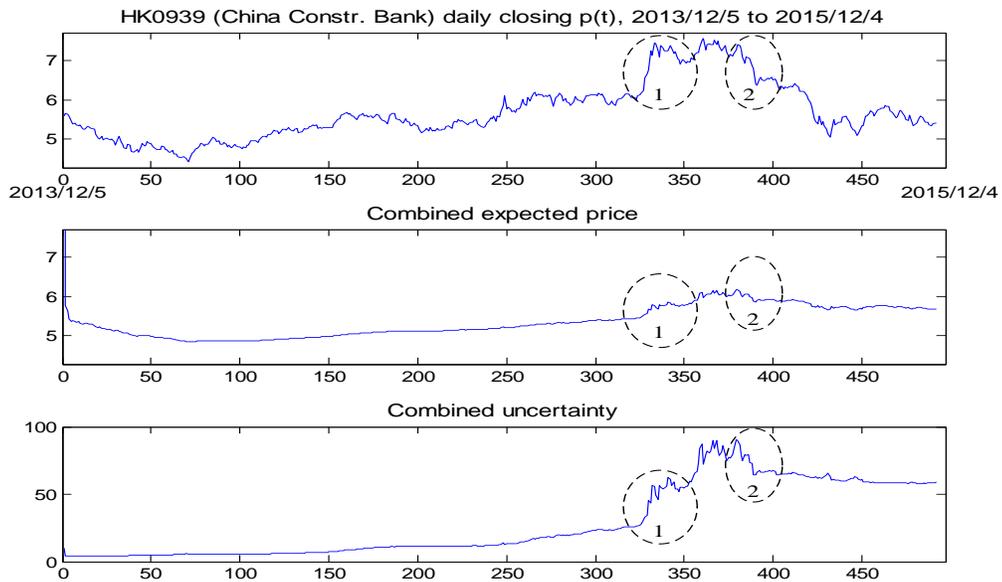

Fig. 24: Same as Fig. 23 for HK0939 (China Constr. Bank). Circle 1: A rapid increase of the combined uncertainty indicates the disagreement of the investors with the new high of the price in Circle 1 so that the price declines sharply after a short period. Circle 2: The clear decrease of the combined uncertainty indicates that the investors agree with the decline of the price in Circle 2 so that the trend continues.

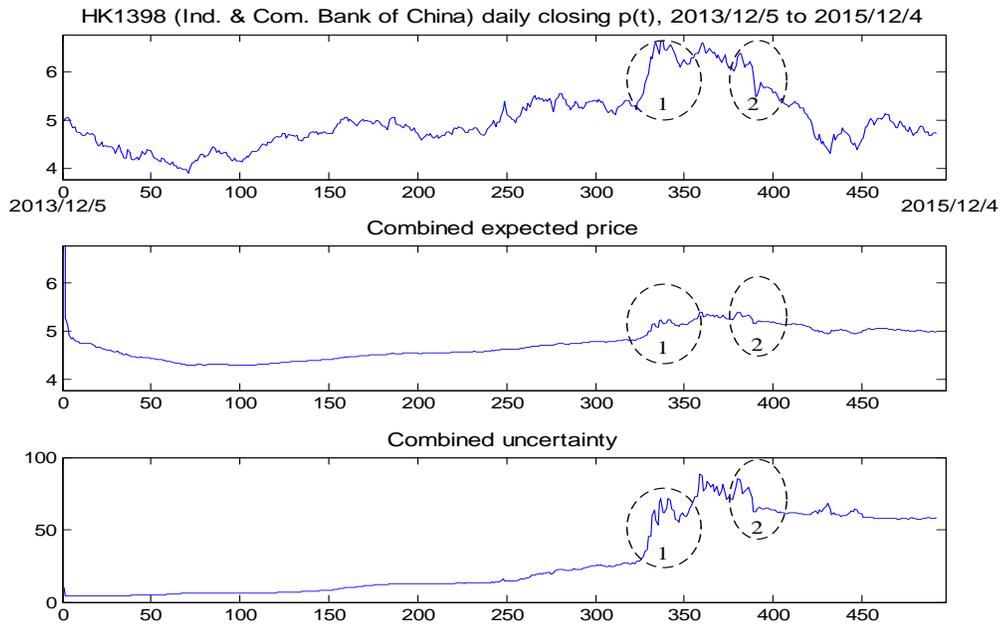

Fig. 25: Same as Fig. 23 for HK1398 (Ind. & Com. Bank of China) and same interpretation for Circles 1 and 2 as for Fig. 24.

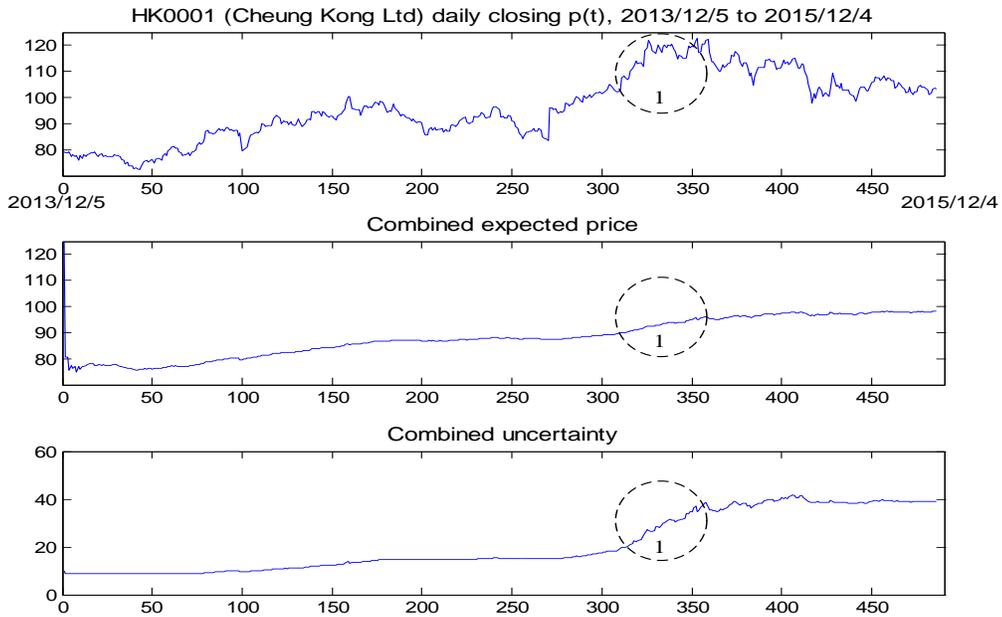

Fig. 26: Same as Fig. 23 for HK0001 (Cheung Kong Ltd). Circle 1: A big increase of the combined uncertainty indicates the disagreement of the investors with the new high of the price in Circle 1 so that a price down trend follows afterwards.

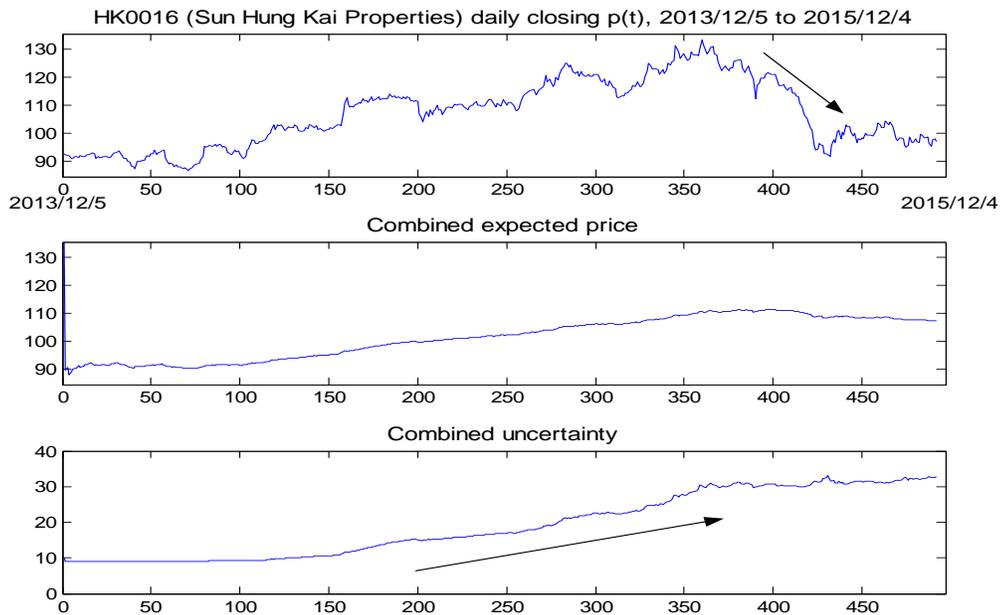

Fig. 27: Same as Fig. 23 for HK0016 (Sun Hung Kai Properties). A steady increase of the combined uncertainty indicates that the investors get more and more uncertain about the uptrend of the price so that a sharp decline is expected.

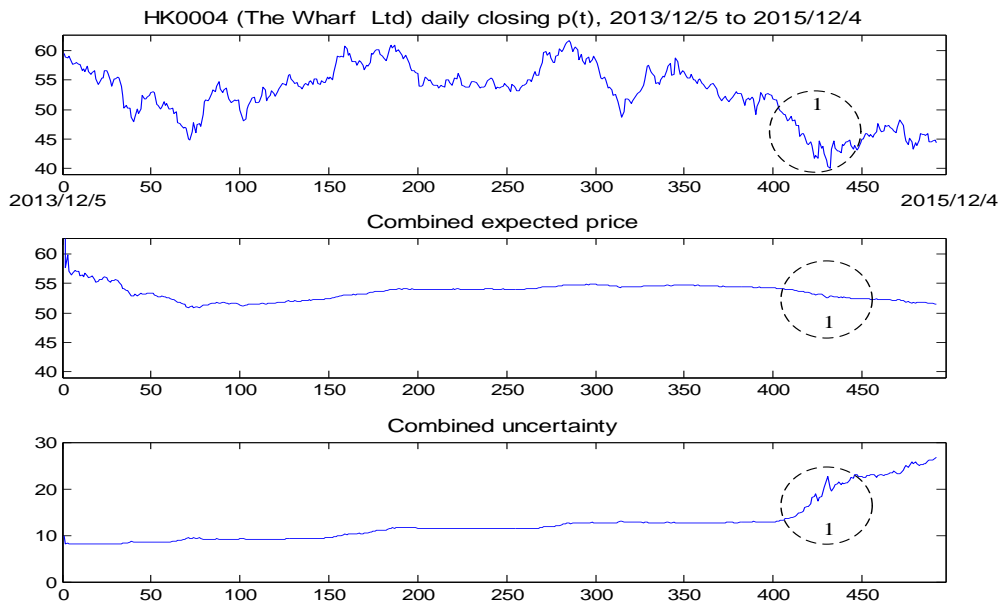

Fig. 28: Same as Fig. 23 for HK0004 (The Wharf Ltd). Circle 1: A big increase of the combined uncertainty indicates the disagreement of the investors with the price decline in Circle 1 so that a price rebound follows afterwards.

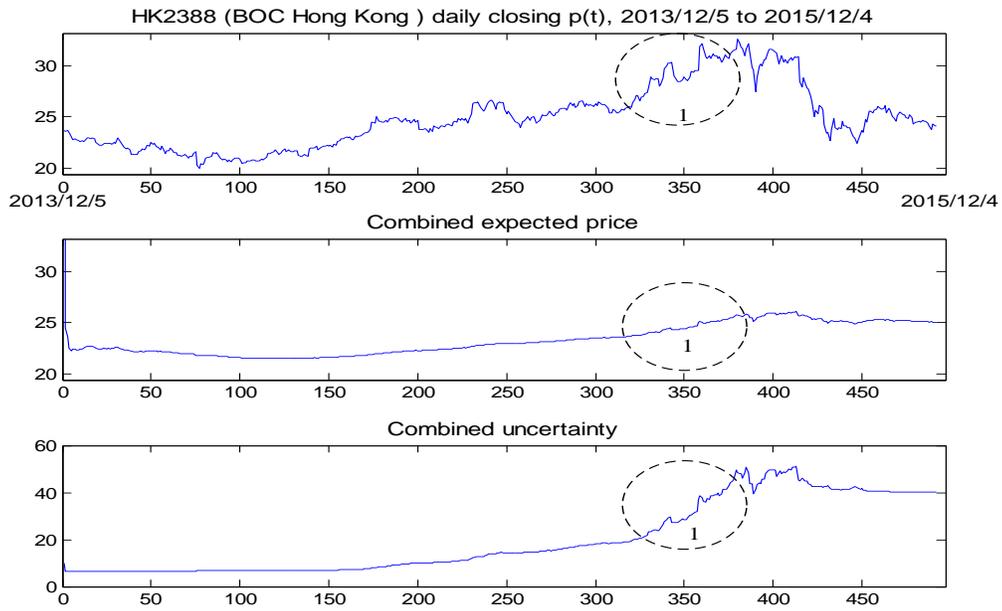

Fig. 29: Same as Fig. 23 for HK2388 (BOC Hong Kong). Circle 1: A steady big increase of the combined uncertainty indicates the increasing disagreement of the investors with the big price rise in Circle 1 so that a sharp price decline is expected in the near future.

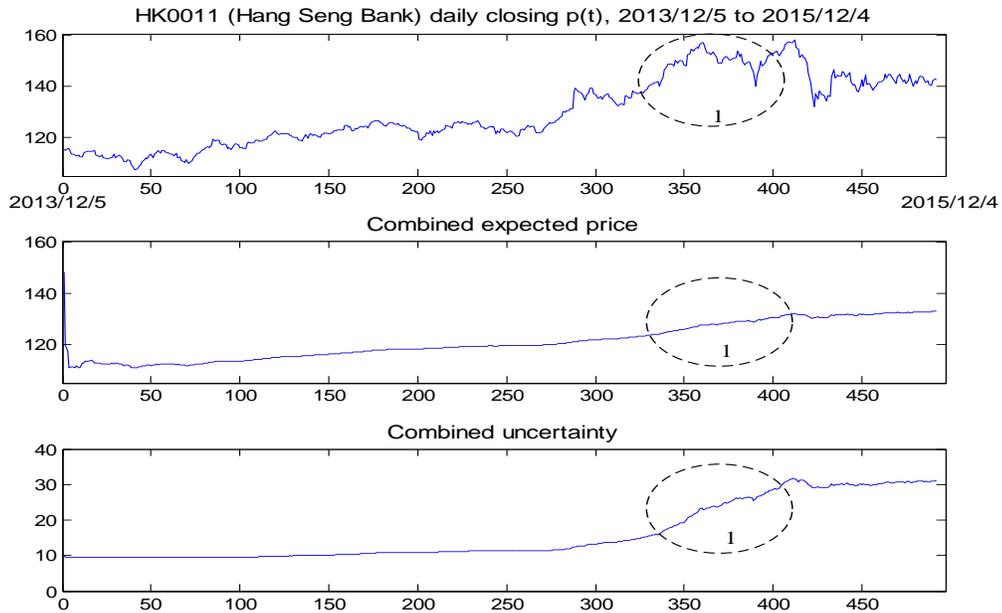

Fig. 30: Same as Fig. 23 for HK0011 (Hang Seng Bank) and the same interpretation for Circle 1 as for Fig. 29.

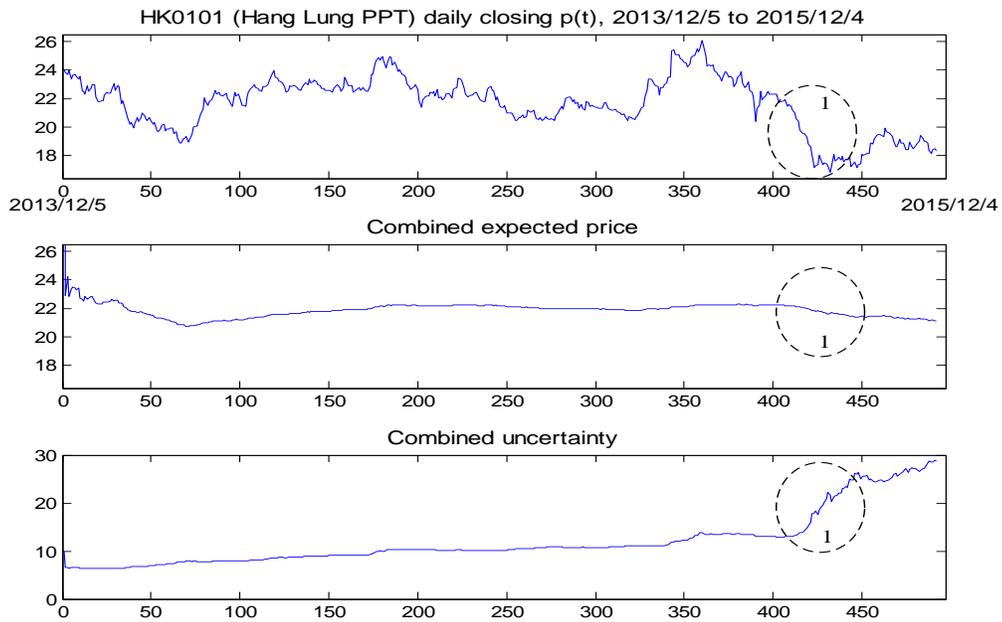

Fig. 31: Same as Fig. 23 for HK0101 (Hang Lung PPT). Circle 1: A big increase of the combined uncertainty indicates the disagreement of the investors with the price decline in Circle 1 so that a price rebound follows afterwards.

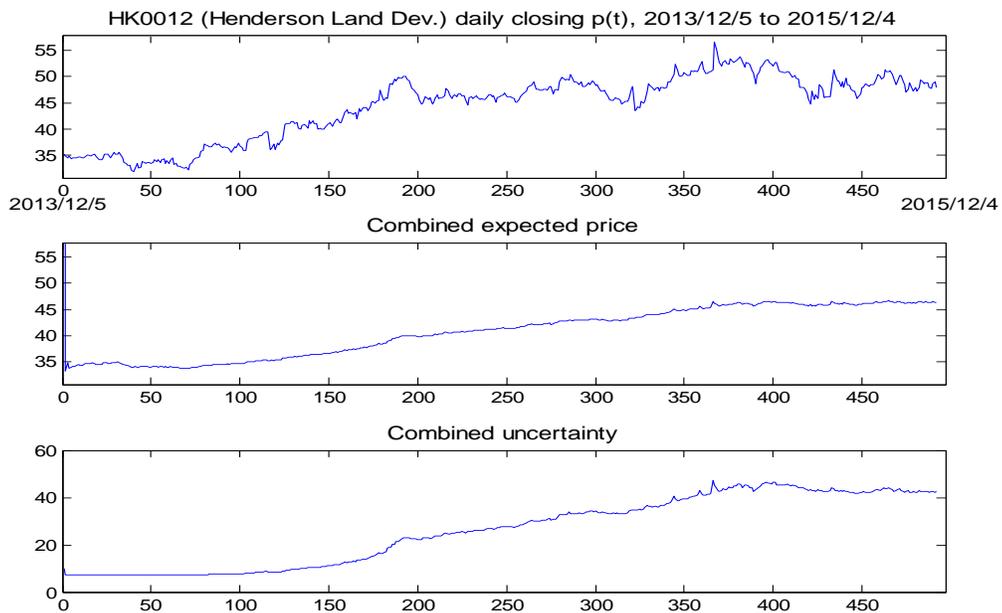

Fig. 32: Same as Fig. 23 for HK0012 (Henderson Land Dev.). A steady increase of the combined uncertainty indicates that the investors get more and more uncertain about the uptrend of the price so that a decline is expected which has not happened yet.

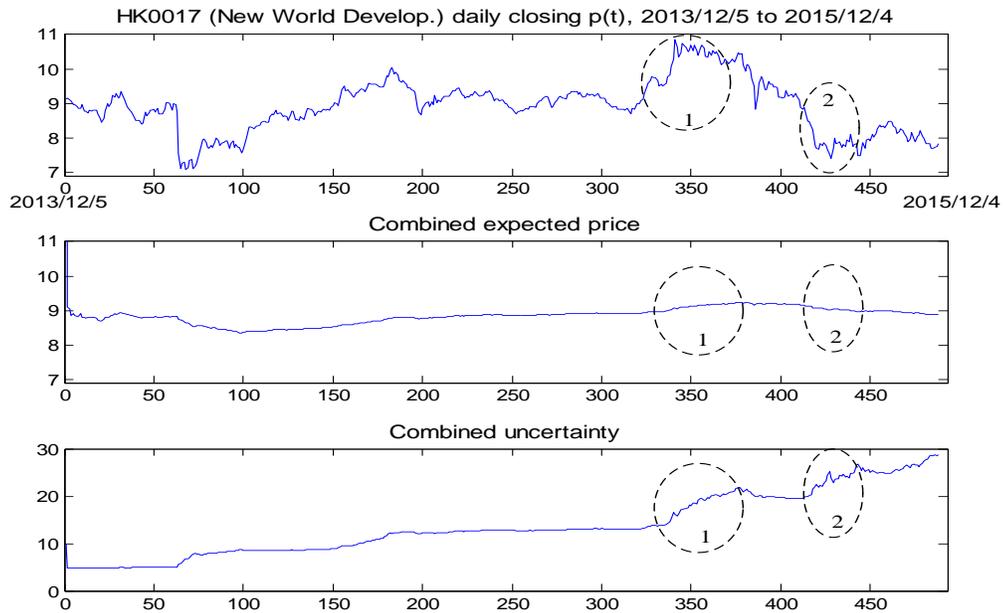

Fig. 33: Same as Fig. 23 for HK0017 (New World Develop.). Circles 1 and 2: Big increases of the combined uncertainty indicate the disagreement of the investors with the sharp price rise in Circle 1 and big drop in Circle 2 such that the price declines after Circle 1 and bounds back after Circle 2.

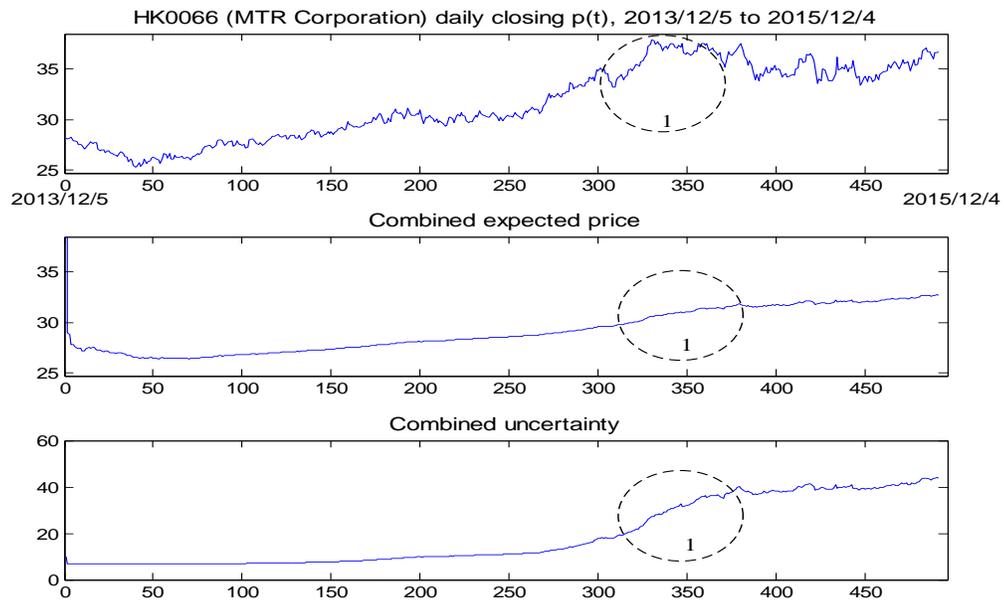

Fig. 34: Same as Fig. 23 for HK0066 (MTR Corporation). Circle 1: A big increase of the combined uncertainty together with a small increase of the combined expected price indicates that the price rise is over and will move horizontally after Circle 1.

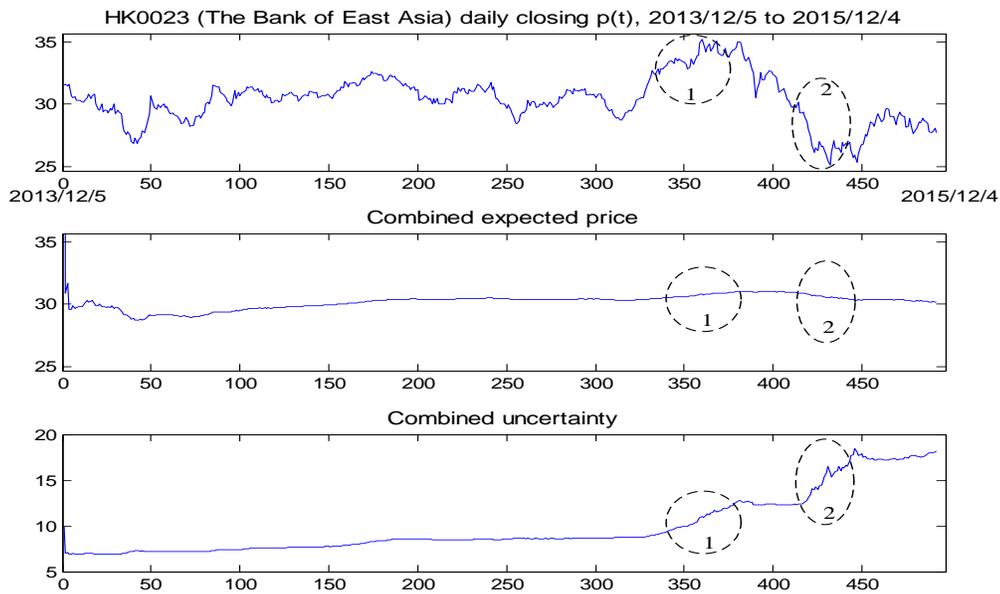

Fig. 35: Same as Fig. 23 for HK0023 (The Bank of East Asia) and the same interpretation for Circles 1 and 2 as for Fig. 33, with a slight difference that a sharper rise of the combined uncertainty in Circle 2 results in a stronger rebound of the price after Circle 2.

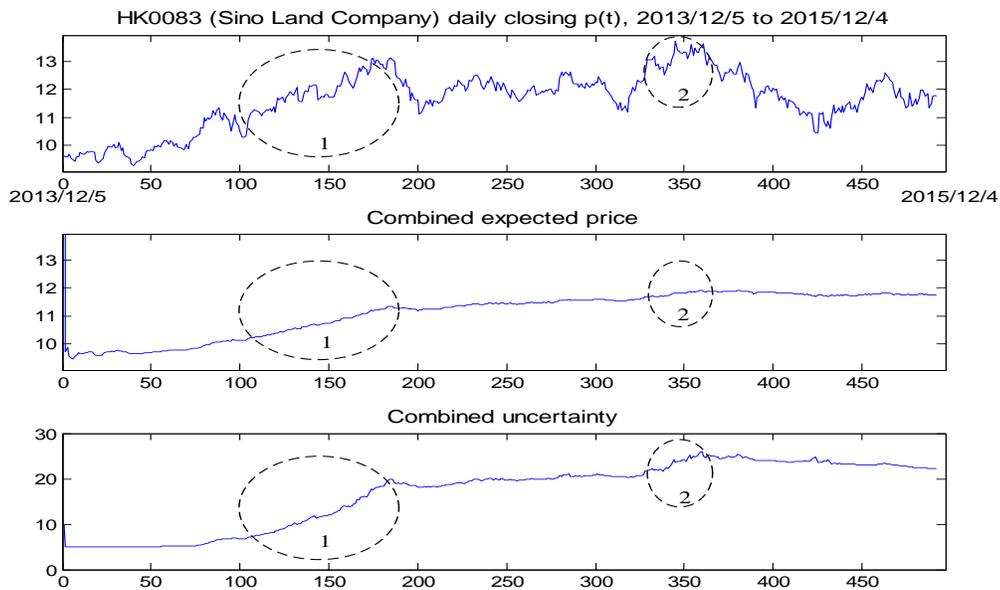

Fig. 36: Same as Fig. 23 for HK0083 (Sino Land Company). Circle 1: A big increase of the combined uncertainty together with a small increase of the combined expected price indicates that the price rise is over and will move horizontally after Circle 1. Circle 2: An increase of the combined uncertainty indicates the disagreement of the investors with the price rise in Circle 2 so that the price moves downwards after Circle 2.

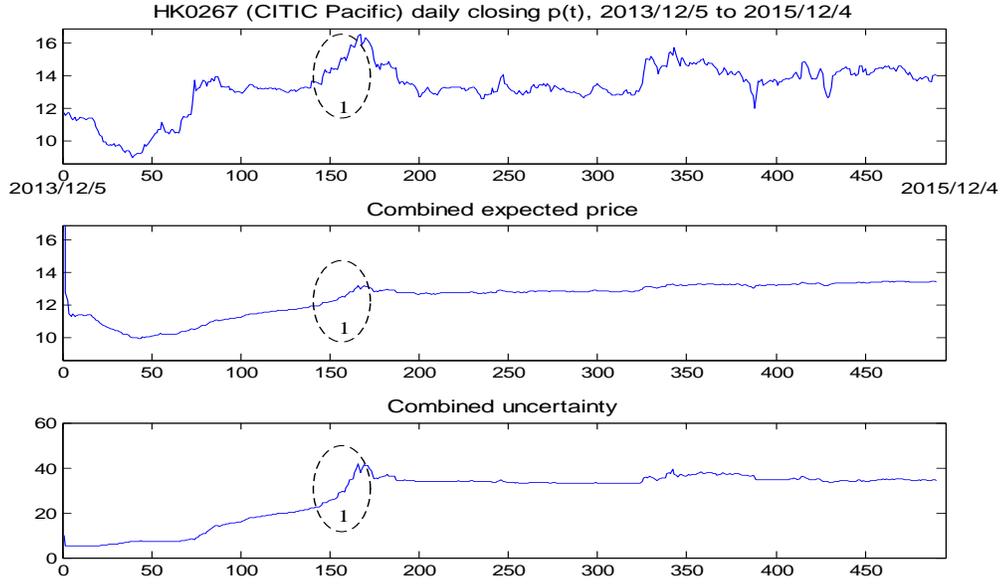

Fig. 37: Same as Fig. 23 for HK0267 (CITIC Pacific). Circle 1: A rapid increase of the combined uncertainty indicates the disagreement of the investors with the sharp rise of the price in Circle 1 so that the price declines after Circle 1.

Observation 2 is important because it shows that the estimated combined uncertainties do have some predicting power for prices in the near future so that some useful trading strategies may be developed. To develop such trading strategies, we first translate Observation 2 into a number of fuzzy IF-THEN rules. Specifically, let $r_p(t) = \ln(p_t/p_{t-\tau})$ and $r_\sigma(t) = \ln(\hat{\sigma}_t/\hat{\sigma}_{t-\tau})$ be the relative changes of the price $p_t$ and the estimated combined uncertainty $\hat{\sigma}_t$, respectively, for a short period of $\tau$ trading days (e.g. $\tau = 5$ means a week), we get the following fuzzy IF-THEN rules from Observation 2:

*Rule 1*: IF $r_p(t)$ is Positive Big and $r_\sigma(t)$ is Positive Big, THEN $p_t$ will decline shortly.

*Rule 2*: IF $r_p(t)$ is Negative Big and $r_\sigma(t)$ is Positive Big, THEN $p_t$ will rise shortly.

*Rule 3*: IF $r_p(t)$ is Positive and $r_\sigma(t)$ is Negative, THEN $p_t$ will continue to rise.

*Rule 4*: IF $r_p(t)$ is Negative and $r_\sigma(t)$ is Negative, THEN $p_t$ will continue to decline.

where Positive Big (PB), Negative Big (NB), Positive (P) and Negative (N) are fuzzy sets with appropriate membership functions $\mu_{PB}, \mu_{NB}, \mu_P$ and $\mu_N$ (such as those in [51]). Then, Rules 1 and 4 lead to the following trading rule:

*Trading Rule 1*: IF $\mu_{PB}\big(r_p(t)\big)\mu_{PB}\big(r_\sigma(t)\big) > \Delta$ or $\mu_N\big(r_p(t)\big)\mu_N\big(r_\sigma(t)\big) > \Delta$,

THEN sell (short) the stock.

and Rules 2 and 3 lead to:

*Trading Rule 2*: IF $\mu_{NB}(r_p(t))\mu_{PB}(r_\sigma(t)) > \Delta$ or $\mu_P(r_p(t))\mu_N(r_\sigma(t)) > \Delta$,

THEN buy (long) the stock.

where $\Delta$ is a threshold. The detailed analyses and testing of these trading rules (and more) will be conducted in another paper.

Finally, it is interesting to estimate the proportion of the market which is taken by investors whose trading decisions are influenced by word-of-mouth. This is a difficult problem if the only information we have is the price data $\{p_t: t = 0,1,2,...\}$; we have to rely on other more detailed information about the investors, such as the membership codes of the brokers which are possible to obtain in some Stock Exchanges (we have no access to these data), to get a more reliable estimate of this proportion. With this said however, it is possible to get a rough estimate of this proportion based on the price data $\{p_t: t = 0,1,2,...\}$ using a price dynamic model. Specifically, suppose we use the price model (62) with $\bar{p}_t$ and $\sigma_t$ provided by the algorithm (65)-(69), then an estimate of the proportion of the market taken by the word-of-mouth investors based on the information set $\{p_t: t = 0,1,2,...,N\}$ may be computed as follows:

$$\frac{word-of-mouth}{market} = \frac{\sum_{t=0}^{N-1}\left|\frac{\ln(\widehat{\bar{p}}_t)-\ln(p_t)}{\hat{\sigma}_t}\right|}{\sum_{t=0}^{N-1}\left|\frac{\ln(\widehat{\bar{p}}_t)-\ln(p_t)}{\hat{\sigma}_t}\right|+\sum_{t=0}^{N-1}|\varepsilon_t|}$$

$$= \frac{\sum_{t=0}^{N-1}\left|\frac{\ln(\widehat{\bar{p}}_t)-\ln(p_t)}{\hat{\sigma}_t}\right|}{\sum_{t=0}^{N-1}\left|\frac{\ln(\widehat{\bar{p}}_t)-\ln(p_t)}{\hat{\sigma}_t}\right|+\sum_{t=0}^{N-1}\left|\ln(p_{t+1})-\ln(p_t)-\frac{\ln(\widehat{\bar{p}}_t)-\ln(p_t)}{\hat{\sigma}_t}\right|} \quad (70)$$

For the fifteen stocks in Figs. 23-37 of the 492 trading days from Dec. 5, 2013 to Dec. 4, 2015, this $\frac{word-of-mouth}{market}$ equals 33.38% (HSBC), 29.87% (China Constr. Bank), 29.53% (Ind. & Com. Bank of China), 28.35% (Cheung Kong Ltd), 31.06% (Sun Hung Kai Properties), 27.59% (The Wharf Ltd), 30.43% (BOC Hong Kong), 33.76% (Hang Seng Bank), 30.92% (Hang Lung PPT), 26.63% (Henderson Land Dev.), 30.82% (New World Develop.), 34.14% (MTR Corporation), 32.14% (The Bank of East Asia), 27.06% (Sino Land Company) and 25.33% (CITIC Pacific), respectively. Of course, these numbers should not be over-interpreted because they are based on the price model (62) which may or may not be a good model for reality.

## VII. Concluding Remarks

Investors are human beings, who live in communities, where they discuss the markets and form opinions about the prices, which determine their investment decisions (buy and sell: what, when, and how many), that in turn drive the price dynamics. A precise and detailed mathematical study of this interplay between investor networks and price dynamics is clearly an important problem, which is largely missing in the literature. The stock price dynamic models proposed in this paper for this interplay are straightforward in concept and easy to understand: The price dynamic equation is a simple excess demand driven equation where the excess demand is propositional to the relative differences between the expected prices of the investors and the real stock price and inversely propositional to the uncertainties about these expected prices, and the expected prices and their uncertainties come from an investor social network called Bounded Confidence Fuzzy Opinion Network (BCFON) where only those investors whose fuzzy expectations about the stock prices are close enough to each other are connected. A key character of the models of this paper is that the investors' opinions are modeled as fuzzy sets so that the interactions between the opinions and their uncertainties can be studied rigorously in a precise mathematical framework. Through detailed mathematical analyses, extensive simulations and thoughtful applications to real stock price data, we reveal the important role that the uncertainties play to shape the price dynamics and the investor social networks. In particular, we discovered that a sharp increase of the combined uncertainty was a reliable signal to predict the reversal of the current price trend when we applied the model to fifteen top banking and real estate stocks in Hong Kong for the recent two-year data from Dec. 5, 2013 to Dec. 4, 2015.

The basic argument of the fuzzy opinion network approach of this paper is that human opinions are inherently fuzzy (uncertain) so that an investor's expected price and the uncertainty about this expected price are two sides of the same coin and therefore should be considered simultaneously when we study the formation and evolution of investors' expectations about future prices. In our Bounded Confidence Fuzzy Opinion Networks, we consider three typical scenarios where investors form their uncertainties about the expected prices: (a) Local Reference, where an investor views the average of his neighbors' opinions as the reference, (b) Global Reference, where an investor views the average of all the investors' (the general public) opinions as the reference, and (c) Real Price Reference, where an investor views the current real price as the reference. We proved that: i) the Local Reference leads to the convergence of the investors into different groups where the investors in the same group share the same expected price with the same uncertainty while the converged expected prices and their uncertainties of different groups are different in general; and ii) the Global Reference and the Real Price Reference lead to the convergence of all the investors into the same expected price with the same uncertainty, where the converged expected price and its uncertainty are constants in the Global Reference case while in the Real Price Reference case they are changing with time. More applications of these three schemes to characterize investors' uncertainties in different situations will be studied in future research.

## Appendix

**Proof of Lemma 2:** Substituting (20) and (21) into the Compositional Rule of Inference (9), we have

$$\mu_X(x) = \max_{v \in R} \min\left[e^{-\frac{|x-v|^2}{\sigma_X^2}}, e^{-\frac{|v-c_V|^2}{\sigma_V^2}}\right] \tag{A1}$$

where the max is achieved at the intersection point:

$$\frac{x-v}{\sigma_X} = \frac{v-c_V}{\sigma_V} \tag{A2}$$

which gives

$$v = \frac{c_V \sigma_X + x \sigma_V}{\sigma_X + \sigma_V} \tag{A3}$$

Substituting (A3) into (A1) yields (22). ∎

**Proof of Theorem 1:** We have already shown in (23) that the dynamic equations (25) and (26) are true for $t = 1$. Now viewing $X_i(1)$ as the initial states and repeating the process, we get (25) and (26) for $t = 2$, and continue to get the general results. (27)-(31) come directly from the definition (11)-(15). ∎

**Proof of Theorem 2:** The proof of Theorem 2 is based on the following result from [36]:

**Lemma 3:** If a row-stochastic matrix $W(t) = [w_{ij}(t)]_{n \times n}$ (all $w_{ij}(t)$ are non-negative and the sum of row equals one: $\sum_{j=1}^{n} w_{ij}(t) = 1$ for all $i = 1, \ldots, n$) satisfies the following three conditions:

i) the diagonal of $W(t)$ is positive, i.e., $w_{ii}(t) > 0$ for $i = 1, \ldots, n$,

ii) the zero-entries in $W(t)$ are symmetric, i.e., for every two investors $i, j \in \{1, \ldots, n\}$ it holds that $w_{ij}(t) > 0 \Leftrightarrow w_{ji}(t) > 0$, and

iii) there is $\delta > 0$ such that the lowest positive entry of $W(t)$ is greater than $\delta$,

then there exists a time step $t_N$ and pairwise disjoint classes of agents $I_1 \cup \cdots \cup I_q = \{1, \ldots, n\}$ such that[3]

$$\lim_{t \to \infty} W(t, 0) = \begin{bmatrix} K_1 & & 0 \\ & \ddots & \\ 0 & & K_q \end{bmatrix} W(t_N, 0) \tag{A4}$$

where $W(t, 0) \equiv W(t-1)W(t-2) \cdots W(0)$ and $K_1, \ldots, K_q$ are square row-stochastic matrices with equal rows in the sizes of $I_1, \ldots, I_q$, respectively (The block structure is achieved by sorting the matrix indices according to $I_1, \ldots, I_q$). ∎

We now prove Theorem 2 based on Lemma 3. Let $\text{Cen}(X(t)) = (\text{Cen}(X_1(t)), \ldots, \text{Cen}(X_n(t)))^T$, $\text{Sdv}(X(t)) = (\text{Sdv}(X_1(t)), \ldots, \text{Sdv}(X_n(t)))^T$, $W(t) = [w_{ij}(t)]_{n \times n}$ and $u(t) = (u_1(t), \ldots, u_n(t))^T$ with $w_{ij}(t)$ given by (38), (39) and $u_i(t)$ given by (40) (Local Reference), then the dynamic equations (25) and (26) can be rewritten in the matrix form:

$$\text{Cen}(X(t+1)) = W(t) \text{Cen}(X(t)) \tag{A5}$$

$$\text{Sdv}(X(t+1)) = W(t) \text{Sdv}(X(t)) + u(t+1) \tag{A6}$$

Solving (A5) we obtain:

$$\text{Cen}(X(t)) = W(t-1)W(t-2) \cdots W(0) \text{Cen}(X(0)) \tag{A7}$$

To apply Lemma 3 to the matrix $W(t) = [w_{ij}(t)]_{n \times n}$ with $w_{ij}(t)$ given by (27), (28), first notice that the $W(t)$ is a row-stochastic matrix because all $w_{ij}(t)$ are non-negative according the definition (27) and $\sum_{j=1}^{n} w_{ij}(t) = \sum_{j \in N_i(t)} \frac{1}{|N_i(t)|} = \frac{|N_i(t)|}{|N_i(t)|} = 1$ for all $i = 1, \ldots, n$. Next, we check that the matrix $W(t)$ satisfies the conditions i)-iii) in Lemma 3. Since $i \in N_i(t)$ according to (39) (notice that $d_i < 1$), we have $|N_i(t)| \geq 1$ and $w_{ii}(t) = \frac{1}{|N_i(t)|} > 0$, therefore condition i) of Lemma 3 is satisfied for the $W(t) = [w_{ij}(t)]_{n \times n}$ with $w_{ij}(t)$ given by (27), (28). Since $d_i = d \in (0,1)$ for all $i = 1, \ldots, n$, it follows from (28) that $j \in N_i(t) \left( e^{-\frac{|\text{Cen}(X_i(t)) - \text{Cen}(X_j(t))|^2}{(\text{Sdv}(X_i(t)) + \text{Sdv}(X_j(t)))^2}} > d \right)$ implies $i \in N_j(t)$

$\left( e^{-\frac{|\text{Cen}(X_j(t)) - \text{Cen}(X_i(t))|^2}{(\text{Sdv}(X_i(t)) + \text{Sdv}(X_j(t)))^2}} > d \right)$ for any $i, j = 1, \ldots, n$, thus condition ii) of Lemma 3 is satisfied for the $W(t)$. Since $|N_i(t)| \leq n$, it follows that any positive $w_{ij}(t) = \frac{1}{|N_i(t)|} > \frac{1}{n+1} \equiv \delta$, hence condition iii) of Lemma 3 is satisfied for the $W(t)$. Now applying the result (A4) of Lemma 3 to our equation (A7), we have

$$\lim_{t \to \infty} \text{Cen}(X(t)) = \begin{bmatrix} K_1 & & 0 \\ & \ddots & \\ 0 & & K_q \end{bmatrix} W(t_N, 0) \text{Cen}(X(0)) \tag{A8}$$

Since $W(t_N, 0)\text{Cen}(X(0))$ is a fixed column vector and $K_1, \ldots, K_q$ are square row-stochastic matrices with equal rows in the sizes of $I_1, \ldots, I_q$, respectively, we have that $\lim_{t \to \infty} \text{Cen}(X_i(t)) = x_{k\infty}$ for $i \in I_k$, $k = 1, \ldots, q$; this proves (32). To prove the finite step convergence of (32), first note that (A8) implies $W(t) = \begin{bmatrix} K_1 & & 0 \\ & \ddots & \\ 0 & & K_q \end{bmatrix}$ for $t \geq t_N$.

Since $K_1, \ldots, K_q$ are square row-stochastic matrices with equal rows, we have
$\text{Cen}(X(t_N + 1)) = W(t_N)\text{Cen}(X(t_N)) = \begin{bmatrix} K_1 & & 0 \\ & \ddots & \\ 0 & & K_q \end{bmatrix} W(t_N, 0)\text{Cen}(X(0)) = \lim_{t \to \infty} \text{Cen}(X(t))$, meaning that all $\text{Cen}(X_i(t))$ reach their final values $x_{k\infty}$ ($k = 1, \ldots, q; i \in I_k$) at $t = t_N + 1$. For $t > t_N + 1$, we have

$$\operatorname{Cen}(X(t+1)) = \begin{bmatrix} K_1 & & 0 \\ & \ddots & \\ 0 & & K_q \end{bmatrix}^{t-t_N-1} \operatorname{Cen}(X(t_N+1)) = \begin{bmatrix} K_1 & & 0 \\ & \ddots & \\ 0 & & K_q \end{bmatrix} \operatorname{Cen}(X(t_N+1)) = \lim_{t \to \infty} \operatorname{Cen}(X(t))$$,

which proves the finite step convergence of $\operatorname{Cen}(X_i(t))$.

To prove the finite step convergence of the uncertainties $\operatorname{Sdv}(X_i(t))$, note that the finite step convergence of $\operatorname{Cen}(X(t))$ implies $u_i(t+1) = b \left| \operatorname{Cen}(X_i(t)) - \frac{1}{|N_i(t)|} \sum_{j \in N_i(t)} \left( \operatorname{Cen}(X_j(t)) \right) \right| = b \left| x_{k\infty} - \frac{1}{|N_i(t)|} \sum_{j \in N_i(t)} (x_{k\infty}) \right| = b |x_{k\infty} - x_{k\infty}| = 0$ for all $t > t_N$, therefore the dynamic equation (A6) for $\operatorname{Sdv}(X(t))$ is reduced to

$$\operatorname{Sdv}(X(t+1)) = W(t) \operatorname{Sdv}(X(t))$$

$$= \begin{bmatrix} K_1 & & 0 \\ & \ddots & \\ 0 & & K_q \end{bmatrix} \operatorname{Sdv}(X(t)) \quad (A9)$$

for $t > t_N$, where $W(t) = \begin{bmatrix} K_1 & & 0 \\ & \ddots & \\ 0 & & K_q \end{bmatrix}$ for $t \geq t_N$ follows from (A8). Since $K_1, \ldots, K_q$ are square row-stochastic matrices with equal rows in the sizes of $I_1, \ldots, I_q$, respectively, setting $t = t_N + 1$ in (A9) we have $\operatorname{Sdv}(X_i(t_N + 2)) = \sigma_{k\infty}$ for $i \in I_k$, $k = 1, \ldots, q$, or in matrix form $\operatorname{Sdv}(X(t_N + 2)) = \left[ \underbrace{\sigma_{1\infty}, \ldots, \sigma_{1\infty}}_{\#I_1}, \cdots, \underbrace{\sigma_{q\infty}, \ldots, \sigma_{q\infty}}_{\#I_q} \right]^T$ where $\#I_k$ ($k = 1, \ldots, q$) denotes the size of $I_k$. Setting $t = t_N + 2$ in (A9) and substituting $\operatorname{Sdv}(X(t_N + 2)) = \left[ \underbrace{\sigma_{1\infty}, \ldots, \sigma_{1\infty}}_{\#I_1}, \cdots, \underbrace{\sigma_{q\infty}, \ldots, \sigma_{q\infty}}_{\#I_q} \right]^T$ into (A9), we have

$$\operatorname{Sdv}(X(t_N + 3)) = \begin{bmatrix} K_1 & & 0 \\ & \ddots & \\ 0 & & K_q \end{bmatrix} \begin{pmatrix} \left.\begin{matrix} \sigma_{1\infty} \\ \vdots \\ \sigma_{1\infty} \end{matrix}\right\} \#I_1 \\ \vdots \\ \left.\begin{matrix} \sigma_{q\infty} \\ \vdots \\ \sigma_{q\infty} \end{matrix}\right\} \#I_q \end{pmatrix}$$

$$= \left[ \underbrace{\sigma_{1\infty}, \ldots, \sigma_{1\infty}}_{\#I_1}, \cdots, \underbrace{\sigma_{q\infty}, \ldots, \sigma_{q\infty}}_{\#I_q} \right]^T \quad (A10)$$

and continuing with this we get $\text{Sdv}(X(t+1)) = \left[ \underbrace{\sigma_{1\infty}, \ldots, \sigma_{1\infty}}_{\#I_1}, \cdots, \underbrace{\sigma_{q\infty}, \ldots, \sigma_{q\infty}}_{\#I_q} \right]^T$ for all $t > t_N$. ∎

**Proof of Theorem 3:** From the proof of Theorem 2 we see that the finite step convergence of $\text{Cen}(X_i(t))$ does not depend on the choice of $u_i(t+1)$, hence the convergence result for $\text{Cen}(X_i(t))$ in Theorem 2 applies to this Global Reference case; that is, we have from Theorem 2 that there exists $t_N$ such that $\text{Cen}(X_i(t)) = x_{k\infty}$, $i \in I_k$, $k = 1, \ldots, q$, for all $t > t_N$. Our task is to prove that the $q$ must be equal to 1, which gives (34). We prove this by contradiction. Assume that $q \geq 2$, then there are $x_{1\infty} \neq x_{2\infty}$ such that when $t > t_N$, $\text{Cen}(X_i(t)) = x_{1\infty}$ for $i \in I_1$ and $\text{Cen}(X_i(t)) = x_{2\infty}$ for $i \in I_2$. Since $x_{1\infty} \neq x_{2\infty}$, we have either $x_{1\infty} \neq \frac{1}{n}\sum_{j=1}^{n} \text{Cen}(X_j(t))$ or $x_{2\infty} \neq \frac{1}{n}\sum_{j=1}^{n} \text{Cen}(X_j(t))$ for $t > t_N$, and we suppose the first inequality is true; then for $i \in I_1$ and $t > t_N$, the Global Reference (30) gives $u_i(t+1) = b\left|x_{1\infty} - \frac{1}{n}\sum_{j=1}^{n} \text{Cen}(X_j(t))\right|$ $= b\left|x_{1\infty} - \frac{1}{q}\sum_{k=1}^{q}(\#I_k)x_{k\infty}\right|$ which is a positive constant. Since $W(t) = \begin{bmatrix} K_1 & & 0 \\ & \ddots & \\ 0 & & K_q \end{bmatrix}$ for $t \geq t_N$ where $K_k = \begin{bmatrix} \frac{1}{\#I_k} & \cdots & \frac{1}{\#I_k} \\ \vdots & \ddots & \vdots \\ \frac{1}{\#I_k} & \cdots & \frac{1}{\#I_k} \end{bmatrix}_{\#I_k \times \#I_k}$ from the definition of $w_{ij}(t)$ in (27), (28), we have from (26) that for $i \in I_1$ and $t > t_N$:

$$\sum_{i \in I_1} \text{Sdv}(X_i(t+1)) = \sum_{i \in I_1}\left[\sum_{j=1}^{n} w_{ij}(t) \text{Sdv}(X_j(t)) + u_i(t+1)\right]$$

$$= \sum_{i \in I_1} \sum_{j \in I_1} \frac{1}{\#I_1} \text{Sdv}(X_j(t)) + b\sum_{i \in I_1}\left|x_{1\infty} - \frac{1}{q}\sum_{k=1}^{q}(\#I_k)x_{k\infty}\right|$$

$$= \sum_{i \in I_1} \text{Sdv}(X_i(t)) + (\#I_1)b\left|x_{1\infty} - \frac{1}{q}\sum_{k=1}^{q}(\#I_k)x_{k\infty}\right| \quad (A11)$$

Since $(\#I_1)b\left|x_{1\infty} - \frac{1}{q}\sum_{k=1}^{q}(\#I_k)x_{k\infty}\right|$ is a positive constant, we get from (A11) that $\sum_{i \in I_1} \text{Sdv}(X_i(t)) \to \infty$ as $t \to \infty$. Since all $\text{Sdv}(X_i(t))$'s are non-negative, $\sum_{i \in I_1} \text{Sdv}(X_i(t)) \to \infty$ implies that at least one of the $\text{Sdv}(X_i(t))$'s for $i \in I_1$ must diverge to infinity, so let $\text{Sdv}(X_{i_0}(t)) \to \infty$ for some $i_0 \in I_1$. Then investor $i_0$ must be eventually connected to all the $n$ investors because the distance from investor $i_0$ to an arbitrary investor $j$ $e^{-\frac{|\text{Cen}(X_{i_0}(t)) - \text{Cen}(X_j(t))|^2}{\left(\text{Sdv}(X_{i_0}(t)) + \text{Sdv}(X_j(t))\right)^2}} \to 1$ as $\text{Sdv}(X_{i_0}(t)) \to \infty$, which implies that

$$N_{i_0}(t) = \left\{ j \in \{1, \ldots, n\} \mid e^{-\frac{|\text{Cen}(X_{i_0}(t)) - \text{Cen}(X_j(t))|^2}{\left(\text{Sdv}(X_{i_0}(t)) + \text{Sdv}(X_j(t))\right)^2}} > d_{i_0} \right\} \text{ equals the whole set } \{1, \ldots, n\}$$

as $t \to \infty$ (notice that the constant $d_{i_0} = d < 1$), but this is impossible if $q \geq 2$, given the block-diagonal structure of $\begin{bmatrix} K_1 & & 0 \\ & \ddots & \\ 0 & & K_q \end{bmatrix}$ whose $i_0$'s row is $\left(\frac{1}{n}, \ldots, \frac{1}{n}\right)$ which is possible only when $q = 1$. Thus it must be true that $q = 1$, and (34), (35) and the finite step convergence follow directly from Theorem 2 with $q = 1$. ∎

**Proof of Theorem 4:** Using the same procedure as the proof of Theorem 3, with the global reference value $\frac{1}{n} \sum_{j=1}^{n} \text{Cen}(X_j(t))$ replaced by the external signal $g(t)$, we obtain the finite step convergence of $\text{Cen}(X_i(t))$ to the consensus $x_\infty$; i.e., there exists $t_N$ such that $\text{Cen}(X_i(t)) = x_\infty$ for all $i = 1, \ldots, n$ and all $t > t_N$. To prove $\text{Sdv}(X_i(t+1)) = \sigma_{t+1}$ for all $i = 1, \ldots, n$ and all $t > t_N$, notice that the finite step convergence of $\text{Cen}(X_i(t))$ to the consensus $x_\infty$ implies that all $u_i(t+1) = b|x_\infty - g(t)|$ (for $i = 1, \ldots, n$) are equal to each other and $W(t) = \begin{bmatrix} \frac{1}{n} & \cdots & \frac{1}{n} \\ \vdots & \ddots & \vdots \\ \frac{1}{n} & \cdots & \frac{1}{n} \end{bmatrix}_{n \times n}$ when $t > t_N$, hence

$$\text{Sdv}(X_i(t+1)) = \sum_{j=1}^{n} w_{ij}(t) \text{Sdv}(X_j(t)) + u_i(t+1)$$

$$= \frac{1}{n} \sum_{j=1}^{n} \text{Sdv}(X_j(t)) + b|x_\infty - g(t)| \qquad (A12)$$

for $t > t_N$. Since the right-hand-side of (A12) does not depend on $i$, we can define $\sigma_{t+1} \triangleq \frac{1}{n}\sum_{j=1}^{n} \text{Sdv}(X_j(t)) + b|x_\infty - g(t)| = \text{Sdv}(X_i(t+1))$ and (A12) becomes

$$\sigma_{t+1} = \frac{1}{n}\sum_{j=1}^{n} \sigma_t + b|x_\infty - g(t)|$$
$$= \sigma_t + b|x_\infty - g(t)| \tag{A13}$$

when $t > t_N$, from which we get $\lim_{t\to\infty} \sigma_t = \sigma_{t_N+1} + \sum_{t=t_N+1}^{\infty} b|x_\infty - g(t)|$. ∎

**Proof of Theorem 5:** (a) Taking expectation on both sides of the price dynamic equation (39) and for $t > t_N$ we have

$$E\{\ln(p_{t+1})\} = E\{\ln(p_t)\} + \sum_{i=1}^{n} \frac{a_i[\ln(\bar{p}_{i,\infty}) - E\{\ln(p_t)\}]}{\sigma_{i,\infty}} \tag{A14}$$

Define new variable

$$z_t = E\{\ln(p_t)\} - \frac{\sum_{i=1}^{n}\left(\frac{a_i \ln(\bar{p}_{i,\infty})}{\sigma_{i,\infty}}\right)}{\sum_{i=1}^{n}\left(\frac{a_i}{\sigma_{i,\infty}}\right)} \tag{A15}$$

and in terms of $z_t$ (A14) becomes

$$z_{t+1} = \left(1 - \sum_{i=1}^{n}\left(\frac{a_i}{\sigma_{i,\infty}}\right)\right) z_t \tag{A16}$$

(substituting $z_t$ of (A15) into (A16) yields (A14)) where $t > t_N$. Hence, if $0 < \sum_{i=1}^{n}\left(\frac{a_i}{\sigma_{i,\infty}}\right) < 2$, then $\left|1 - \sum_{i=1}^{n}\left(\frac{a_i}{\sigma_{i,\infty}}\right)\right| < 1$ and (A16) yields $\lim_{t\to\infty} z_t = \left(1 - \sum_{i=1}^{n}\left(\frac{a_i}{\sigma_{i,\infty}}\right)\right)^{\infty} z_{t_N+1} = 0$ which, from the definition of $z_t$ (A15), gives (46).

(b) Since the Global Reference result is a special case of the Local Reference result with $\bar{p}_{i,\infty} = \bar{p}_\infty$ and $\sigma_{i,\infty} = \sigma_\infty$ for all $i = 1, \dots, n$, we get from (46) that $\lim_{t\to\infty} E\{\ln(p_t)\} = \frac{\sum_{i=1}^{n}\left(\frac{a_i \ln(\bar{p}_\infty)}{\sigma_\infty}\right)}{\sum_{i=1}^{n}\left(\frac{a_i}{\sigma_\infty}\right)} = \ln(\bar{p}_\infty)$.

(c) Taking expectation on both sides of (39) and using the results of Theorem 4, we have for $t > t_N$ that

$$E\{\ln(p_{t+1})\} = E\{\ln(p_t)\} + E\left\{\frac{\ln(\bar{p}_\infty) - \ln(p_t)}{\sigma_t}\right\} \sum_{i=1}^{n} a_i \qquad (A17)$$

where $\sigma_t$ follows the equation

$$\sigma_{t+1} = \sigma_t + b|\bar{p}_\infty - p_t| \qquad (A18)$$

which is (37) of Theorem 4 with $x_\infty = \bar{p}_\infty$ and $g(t) = p_t$. Since the noise term $\varepsilon_t$ in the price equation (39) is Gaussian with a constant non-zero standard deviation, $|\bar{p}_\infty - p_t|$ cannot converge to zero as $t \to \infty$ for a typical realization of the random prices $p_t$. Hence, $\sigma_t$ for $t > t_N$ is a non-decreasing sequence diverging to infinity as $t \to \infty$ ($\lim_{t \to \infty} \sigma_t = \sigma_{t_N+1} + \sum_{t=t_N+1}^{\infty} b|\bar{p}_\infty - p_t| = \infty$ if $|\bar{p}_\infty - p_t| \nrightarrow 0$ as $t \to \infty$). We now show $E\left\{\frac{\ln(\bar{p}_\infty) - \ln(p_t)}{\sigma_t}\right\} \to 0$ as $t \to \infty$. For a realization of $p_t$, if $\ln(p_t)$ is bounded, then $\sigma_t \to \infty$ implies $\frac{\ln(\bar{p}_\infty) - \ln(p_t)}{\sigma_t} \to 0$; if $\ln(p_t)$ goes to infinity, then $\sigma_t$ also goes to infinity at a speed no slower than $p_t$ due to (A18), thus $\frac{\ln(\bar{p}_\infty) - \ln(p_t)}{\sigma_t} \to 0$ because $\lim_{x \to \infty} \frac{\ln(x)}{x} = \lim_{x \to \infty} \frac{(\ln(x))'}{x'} = \lim_{x \to \infty} \frac{1}{x} = 0$. That is, for a typical realization of $p_t$ we have $\frac{\ln(\bar{p}_\infty) - \ln(p_t)}{\sigma_t} \to 0$, therefore $E\left\{\frac{\ln(\bar{p}_\infty) - \ln(p_t)}{\sigma_t}\right\} \to 0$ as $t \to \infty$. Now for any positive integer $\tau$ and $t > t_N$, we have from (A17) that

$$|E\{\ln(p_{t+\tau})\} - E\{\ln(p_t)\}| \leq \sum_{i=1}^{n} a_i \sum_{j=0}^{\tau-1} \left|E\left\{\frac{\ln(\bar{p}_\infty) - \ln(p_{t+j})}{\sigma_{t+j}}\right\}\right| \qquad (A19)$$

Since $E\left\{\frac{\ln(\bar{p}_\infty) - \ln(p_{t+j})}{\sigma_{t+j}}\right\} \to 0$, (A19) gives $|E\{\ln(p_{t+\tau})\} - E\{\ln(p_t)\}| \to 0$ as $t \to \infty$, which means that $E\{\ln(p_t)\}$ is a Cauchy sequence and therefore convergent; this proves (48). ∎

**Proof of Theorem 6:** (a) The assumption that any manipulator in $I_0 = \{1, ..., m\}$ never takes other investor as neighbor implies that $N_i(t) = \{i\}$ for $i \in I_0$, thus (56) and (57) are obtained from (40)-(42) and (45).

(b) From (57) we have

$$\lim_{t \to \infty} \sigma_{i,t} = \sigma_{i,0} + \sum_{t=0}^{\infty} b|\bar{p}_{i,0} - p_t| \qquad (A20)$$

for $i \in I_0$. Since $p_t$ does not converge to any constant, we have $|\bar{p}_{i,0} - p_t| \not\to 0$ so that $\lim_{t\to\infty} \sigma_{i,t} = \sigma_{i,0} + \sum_{t=0}^{\infty} b|\bar{p}_{i,0} - p_t| = \infty$ for $i \in I_0$, therefore the neighborhood set

$$N_j(t) = \left\{ i \in \{1, \ldots, n\} \mid e^{-\frac{|\bar{p}_{j,t} - \bar{p}_{i,t}|^2}{(\sigma_{j,t} + \sigma_{i,t})^2}} > d \right\}$$

of any ordinary investor $j \in I_1$ must eventually contain the $m$ manipulators $I_0 = \{1, \ldots, m\}$ as $t \to \infty$ because $e^{-\frac{|\bar{p}_{j,t} - \bar{p}_{i,t}|^2}{(\sigma_{j,t} + \sigma_{i,t})^2}} \to 1$ when $\sigma_{i,t} \to \infty$ and $d < 1$. Hence, from the dynamic equation of the uncertainties $\sigma_{j,t+1} = \frac{1}{|N_j(t)|} \sum_{k \in N_j(t)} \sigma_{k,t} + u_j(t+1)$ we conclude that $\sigma_{j,t}$ for all $j \in I_1 = \{m+1, \ldots, n\}$ must also go to infinity because $\sum_{k \in N_j(t)} \sigma_{k,t}$ contains $\sum_{i \in I_0} \sigma_{i,t}$ which goes to infinity as $t \to \infty$, and this means that $N_j(t) = \left\{ i \in \{1, \ldots, n\} \mid e^{-\frac{|\bar{p}_{j,t} - \bar{p}_{i,t}|^2}{(\sigma_{j,t} + \sigma_{i,t})^2}} > d \right\}$ of any $j \in I_1$ will eventually contain all the $n$ investors as $t \to \infty$. Furthermore, $e^{-\frac{|\bar{p}_{j,t} - \bar{p}_{i,t}|^2}{(\sigma_{j,t} + \sigma_{i,t})^2}} \to 1$ and $d < 1$ ensure that there exists $t_N$ such that $N_j(t) = \{1, \ldots, n\}$ for all $j \in I_1$ when $t > t_N$. Therefore, from (40) we have for $j \in I_1$ and $t > t_N$ that

$$\bar{p}_{j,t+1} = \frac{1}{|N_j(t)|} \sum_{k \in N_j(t)} \bar{p}_{k,t}$$

$$= \frac{1}{n} \sum_{k=1}^{n} \bar{p}_{k,t} \tag{A21}$$

Since $\frac{1}{n} \sum_{k=1}^{n} \bar{p}_{k,t}$ does not depend on $j \in I_1$, we define $\bar{p}_{t+1} \triangleq \frac{1}{n} \sum_{k=1}^{n} \bar{p}_{k,t} = \bar{p}_{j,t+1}$ for $j \in I_1 = \{m+1, \ldots, n\}$ and from (A21) to get

$$\bar{p}_{t+1} = \frac{1}{n} \left( \sum_{i=1}^{m} \bar{p}_{i,t} + \sum_{j=m+1}^{n} \bar{p}_{j,t} \right)$$

$$= \frac{1}{n} \left( \sum_{i=1}^{m} \bar{p}_{i,0} + (n-m)\bar{p}_t \right) \tag{A22}$$

for $t > t_N + 1$. Solving (A22) for $\bar{p}_t$ with initial $\bar{p}_{t_N+2}$, we obtain

$$\bar{p}_t = \left(\frac{n-m}{n}\right)^{t-t_N-2} \bar{p}_{t_N+2} + \left[\left(\frac{n-m}{n}\right)^{t-t_N-3} + \cdots + 1\right] \frac{1}{n} \sum_{i=1}^{m} \bar{p}_{i,0}$$

$$= \left(\frac{n-m}{n}\right)^{t-t_N-2} \bar{p}_{t_N+2} + \frac{1-\left(\frac{n-m}{n}\right)^{t-t_N-3}}{1-\left(\frac{n-m}{n}\right)} \left(\frac{1}{n}\sum_{i=1}^{m} \bar{p}_{i,0}\right) \quad (A23)$$

and

$$\lim_{t \to \infty} \bar{p}_t = \frac{1}{1-\left(\frac{n-m}{n}\right)} \left(\frac{1}{n}\sum_{i=1}^{m} \bar{p}_{i,0}\right)$$

$$= \frac{1}{m}\sum_{i=1}^{m} \bar{p}_{i,0} \quad (A24)$$

which is (58).

To prove the finite step convergence of the uncertainties $\sigma_{j,t}$ of $j \in I_1$ to the same $\sigma_t$ and (59), using the fact that $N_j(t) = \{1, \ldots, n\}$ for all $j \in I_1$ when $t > t_N$, we get

$$\sigma_{j,t+1} = \frac{1}{|N_j(t)|} \sum_{k \in N_j(t)} \sigma_{k,t} + b|\bar{p}_{j,t} - p_t|$$

$$= \frac{1}{n}\sum_{k=1}^{n} \sigma_{k,t} + b|\bar{p}_t - p_t| \quad (A25)$$

for $j \in I_1$ and $t > t_N + 1$. Since $\frac{1}{n}\sum_{k=1}^{n} \sigma_{k,t} + b|\bar{p}_t - p_t|$ does not depend on $j \in I_1$, we define $\sigma_{t+1} \triangleq \frac{1}{n}\sum_{k=1}^{n} \sigma_{k,t} + b|\bar{p}_t - p_t| = \sigma_{j,t+1}$ for $j \in I_1 = \{m+1, \ldots, n\}$ and from (A25) to get

$$\sigma_{t+1} = \frac{1}{n}\left(\sum_{i=1}^{m} \sigma_{i,t} + \sum_{j=m+1}^{n} \sigma_{j,t}\right) + b|\bar{p}_t - p_t|$$

$$= \frac{n-m}{n}\sigma_t + \frac{1}{n}\sum_{i=1}^{m} \sigma_{i,t} + b\left|\frac{1}{m}\sum_{i=1}^{m} \bar{p}_{i,0} - p_t\right| \quad (A26)$$

for $t > t_N + 1$, and (59) is proven. ∎

# References

[1] Acemoglu, D. and A. Ozdaglar, "Opinion Dynamics and Learning in Social Networks," *Dynamic Games and Applications* 1.1: 3-49, 2011.

[2] Åström, K.J. and B. Wittenmark, *Adaptive Control (2$^{nd}$ Edition),* Addison-Wesley Publishing Company, 1995.

[3] Bachelier, L., "Theorie de la speculation." Thesis. Annales Scientifiques de l'Ecole Normale Superieure, 1990. III-17, 21-86. Translated by Cootner (ed.) as "Random character of Stock Market Prices." (MIT, 1964: 17-78).

[4] Bala, V. and S. Goyal, "Learning from Neighbors," *Review of Economic Studies* 65: 595-621, 1998.

[5] Banerjee, A. and D. Fudenberg, "Word-of-mouth learning," *Games and Economics Behavior* 46: 1-22, 2004.

[6] Blondel, V.D., J.M. Hendrickx and J.N. Tsitsiklis, "On Krause's Multi-Agent Consensus Model with State-Dependent Connectivity," *IEEE Trans. on Automatic Control* 54(11): 2586-2597, 2009.

[7] Boero, R., M. Morini, M. Sonnessa and P. Terna, *Agent-Based Models of the Economy: From Theories to Applications*, Palgrave Macmillan, 2015.

[8] Bramoulle, Y., H. Djebbari and B. Fortin, "Identification of peer effects through social networks," *Journal of Econometrics* 150: 41-55, 2009.

[9] Coval, J.D. and T.J. Moskowitz, "Home bias at home: Local equity preference in domestic portfolios," *Journal of Finance* 54(6): 2045-2073, 1999.

[10] Coval, J.D. and T.J. Moskowitz, "The geography of investment: Informed trading and asset prices," *Journal of Political Economy* 109 (4): 811-841, 2001.

[11] Cohen, L., A. Frazzini and C. Malloy, "Sell side school ties," *Journal of Finance* 65: 1409-1437, 2009.

[12] DeGroot, M.H., "Reaching A Consensus," *Journal of the American Statistical Association* 69: 118-121, 1974.

[13] Ellison, G. and D. Fudenberg, "Word-of-mouth communication and social learning," *Quarterly Journal of Economics* 110(1): 93-125, 1995.

[14] Elton, E.J., M.J. Gruber, S.J. Brown and W.N. Goetzmann, *Modern Portfolio Theory and Investment Analysis (7$^{th}$ Edition)*, John Wiley & Sons, Inc., 2007.

[15] Fama, E., "Efficient Capital Marets: A Review of Theory and Empirical Work," *Journal of Finance* 25: 383-417, 1970.

[16] Fama, E., "Two pillars of asset pricing," *American Economic Review* 104(6): 1467-1485, 2014.

[17] Fama, E. and K. French, "Common risk factors in the returns on stocks and bonds," *Journal of Financial Economics* 33: 3-56, 1993.

[18] Fama, E. and K. French, "A five-factor asset pricing model," http://ssrn.com/abstract=2287202, 2014.

[19] Feng, L. and M. Seasholes, "Correlated trading and location," *Journal of Finance* 59(5): 2117-2144, 2004.

[20] Frydman, C., "What Drives Peer Effects in Financial Decision-Making? Neural and Behavioral Evidence," http://ssrn.com/abstract=2561083, 2015.

[21] Ghaderi, J. and R. Srikant, "Opinion Dynamics in Social Networks with Stubborn Agents: Equilibrium and Convergence Rate," *Automatica* 50(12): 3209-3215, 2014.

[22] Golub, B. and M.O. Jackson, "Naïve Learning in Social Networks: Convergence, Influence, and the Wisdom of Crowds," http://www.standford.edu/~jacksonm/naivelearning.pdf, 2007.

[23] Goyal, A. and I. Welch, "A Comprehensive Look at the Empirical Performance of Equity Premium Prediction," *Review of Financial Studies*, 21: 1455-1508, 2008.

[24] Hegselmann, R. and U. Krause, "Opinion Dynamics and Bounded Confidence: Models, Analysis, and Simulations," *Journal of Artificial Societies and Social Simulations* 5(3): http://jasss.soc.surrey.ac.uk/5/3/2.html, 2002.

[25] Hegselmann, R. and U. Krause, "Opinion Dynamics under the Influence of Radical Groups, Charismatic Leaders, and Other Constant Signals: A Simple Unifying Model," *Networks and Heterogeneous Media* 10(3): 477-509, 2015.

[26] Hommes, C.H., "Heterogeneous Agent Models in Economics and Finance," in L. Tesfatsion and K.L. Judd, eds., *Handbook of Computational Economics Vol. 2: Agent-Based Computational Economics*, Elsevier B.V.: 1109-1186, 2006.

[27] Hong, H. and L. Kostovetsky, "Red and blue investing: Values and finance," *Journal of Financial Economics* 103(1): 1-19, 2012.

[28] Hong, H., J.D. Kubik and J.C. Stein, "The neighbor's portfolio: Word-of-mouth effects in the holdings an trades of money managers," *Journal of Finance* 60(6): 2801-2824, 2005.

[29] Horn, R.A. and C.R. Johnson, *Matrix Analysis (2$^{nd}$ Edition)*, Cambridge University Press, 2013.

[30] Hull, J.C., *Options, Futures, and Other Derivatives (7$^{th}$ Edition)*, Pearson Education, Inc., 2009.

[31] Ivkovic, Z. and S. Weisbenner, "Information diffusion effects in individual investors' common stock purchases: Covet the neighbors' investment choices," *Review of Financial Studies* 20(4): 1327-1357, 2007.

[32] Jackson, M.O., *Social and Economic Networks*, Princeton University Press, 2008.

[33] Jensen, F.V., *Bayesian Networks and Decision Graphs (2$^{nd}$ Edition)*, Springer-Verlag, 2007.

[34] Keynes, J.M., *The General Theory of Employment, Interest and Money*, (BN Publishing, 2009), 1935.

[35] Kyle, A.S., "Continuous Auction and Insider Trading," *Econometrica* 53: 1315-1335, 1985.

[36] Lorenz, J., "A Stabilization Theorem for Dynamics of Continuous Opinions," *Physica A* 355: 217-223, 2005.

[37] Lorenz, J.,."Heterogeneous Bounds of Confidence: Meet, Discuss and Find Consensus", *Complexity* 4(15): 43–52, 2010.

[38] Lux, T. and M. Marchesi, "Scaling and Criticality in A Stochastic Multi-Agent Model of A Financial Market," *Nature* 397: 498-500, 1999.

[39] Newman, M., A.L. Barabasi and D.J. Watts, Eds., *The Structure and Dynamics of Networks*, Princeton University Press, Princeton, NJ, 2006.

[40] Olfati-Saber, R., J.A. Fax and R.M. Murray, "Consensus and Cooperation in Networked Multi-Agent Systems," *Proceedings of the IEEE* 95(1): 215-233, 2007.

[41] Pool, V.K., N. Stoffman and S.E. Yonker, "The People in Your Neighborhood: Social Interactions and Mutual Fund Portfolios," *Journal of Finance* 70(6): 2679-2732, 2015.

[42] Sharpe, W., "Capital asset prices: A theory of market equilibrium under conditions of risk," *Journal of Finance* 19: 425-442, 1964.

[43] Scott, J., *Social Network Analysis (3rd Edition)*, Sage Publications, London, 2013.

[44] Shiller, R.J., *Irrational Exuberance*, Broadway Books, New York, 2006.

[45] Shiller, R.J., "Speculative asset prices," *American Economic Review* 104(6): 1486-1517, 2014.

[46] Shiller, R.J. and J. Pound, "Survey evidence on diffusion of interest and information among investors," *Journal of Economic Behavior and Organization* 12(1): 47-66, 1989.

[47] Shiryaev, A.N., *Essentials of Stochastic Finance: Facts, Models, Theory*, World Scientific Publishing Co., 1999.

[48] Sornette, D., "Physics and financial economics (1776-2014): Puzzles, Ising and agent-based models," *Rep. Prog. Phys.* 77, 2014.

[49] Tesfatsion, T. and K.L. Judd, eds., *Handbook of Computational Economics Vol. 2: Agent-Based Computational Economics*, Elsevier B.V., 2006.

[50] Wang, L.X., *A Course in Fuzzy Systems and Control*, Prentice-Hall, NJ, 1997.

[51] Wang, L.X., "Dynamical Models of Stock Prices Based on Technical Trading Rules Part I: The Models; Part II: Analysis of the Model; Part III: Application to Hong Kong Stocks," *IEEE Trans. on Fuzzy Systems* 23(4): 787-801 (Part I); 1127-1141 (Part II); 23(5): 1680-1697 (Part III), 2015.

[52] Wang, L.X. and J. M. Mendel, "Fuzzy opinion networks: A mathematical framework for the evolution of opinions and their uncertainties across social networks," *IEEE Trans. on Fuzzy Systems*, accepted for publication (available under "Early Access"), 2015.

[53] Wasserman, S. and K. Faust, *Social Network Analysis: Methods and Applications*, Cambridge University Press, New York, 1994.

[54] Zadeh, L.A., "Outline of a New Approach to the Analysis of Complex Systems and Decision Processes," *IEEE Trans. on Systems, Man, and Cybern.* 3: 28-44, 1973.

[55] Zadeh, L.A., "The Concept of a Linguistic Variable and Its Application to Approximate Reasoning – I," *Information Sciences*, Vol. 8, pp. 199-249, 1975.

[56] Zadeh, L.A., "Is there a need for fuzzy logic?" *Information Sciences* 178(13): 2751-2779, 2008.